\documentclass[useAMS,usenatbib]{mn2e}
\usepackage{graphicx}
\usepackage{amsmath}
\usepackage{amssymb}
\usepackage{bm,url,times}
\usepackage{float}
\usepackage{xcolor}

\def\Pm{\mbox{P}_M}
\def\Rm{\mbox{R}_M}

\def\urms{{ u_{rms}}}
\def\Brms{{B_{rms}}}

\def\aap{A\&A}
\def\apjl{ApJ}

\def\apjs{ApJS}

\newcommand{\EQ}{\begin{equation}}
\newcommand{\EN}{\end{equation}}
\newcommand{\EQA}{\begin{eqnarray}}
\newcommand{\ENA}{\end{eqnarray}}

\newcommand{\Eq}[1]{Eq.~(\ref{#1})}
\newcommand{\Eqs}[2]{Eqs.~(\ref{#1}) and~(\ref{#2})}

\newcommand{\Fig}[1]{Fig.~\ref{#1}}

\newcommand{\Figs}[2]{Figs.~\ref{#1} and \ref{#2}}

\newcommand{\bra}[1]{\langle #1\rangle}

{}
{}
{}

{}
{}
{}
{}
{}
{}
{}
{}
{}
{}
{}
{}
{}
{}
{}
{}
{}
{}
{}
{}
{}
{}

{}
{}
{}

{}

{}
{}
{}

\newcommand{\xxx}{\hat{\mbox{\boldmath $x$}} {}}
\newcommand{\yyy}{\hat{\mbox{\boldmath $y$}} {}}
\newcommand{\zzz}{\hat{\mbox{\boldmath $z$}} {}}

\newcommand{\ww}{\mbox{\boldmath $w$} {}}

\newcommand{\kk}{\bm{k}}

\newcommand{\xx}{\bm{x}}

\newcommand{\BB}{\bm{B}}

\newcommand{\uu}{\bm{u}}

\newcommand{\JJ}{\mbox{\boldmath $J$} {}}

\newcommand{\AAA}{\mbox{\boldmath $A$} {}}

\newcommand{\FF}{\mbox{\boldmath $F$} {}}

\newcommand{\nab}{\mbox{\boldmath $\nabla$} {}}

\newcommand{\SSSS}{\mbox{\boldmath ${\sf S}$} {}}

\def\la{\mathrel{\mathchoice {\vcenter{\offinterlineskip\halign{\hfil
$\displaystyle##$\hfil\cr<\cr\sim\cr}}}
{\vcenter{\offinterlineskip\halign{\hfil$\textstyle##$\hfil\cr<\cr\sim\cr}}}
{\vcenter{\offinterlineskip\halign{\hfil$\scriptstyle##$\hfil\cr<\cr\sim\cr}}}
{\vcenter{\offinterlineskip\halign{\hfil$\scriptscriptstyle##$\hfil\cr<\cr\sim\cr}}}}}
\def\ga{\mathrel{\mathchoice {\vcenter{\offinterlineskip\halign{\hfil
$\displaystyle##$\hfil\cr>\cr\sim\cr}}}
{\vcenter{\offinterlineskip\halign{\hfil$\textstyle##$\hfil\cr>\cr\sim\cr}}}
{\vcenter{\offinterlineskip\halign{\hfil$\scriptstyle##$\hfil\cr>\cr\sim\cr}}}
{\vcenter{\offinterlineskip\halign{\hfil$\scriptscriptstyle##$\hfil\cr>\cr\sim\cr}}}}}
\def\Pm{\mbox{\rm Pr}_M}
\def\Rm{R_{\rm m}}

\def\EEK{{\cal E}_{\rm K}}
\def\EEM{{\cal E}_{\rm M}}

\def\Brms{B_{\rm rms}}

\def\urms{u_{\rm rms}}

\graphicspath{{./figs/}{./png/}}

\title[]
{Inverse energy transfer in decaying, three dimensional, nonhelical magnetic turbulence due to magnetic reconnection}
\author[]{Pallavi Bhat$^{1,2}$\thanks{
P.Bhat@leeds.ac.uk}, Muni Zhou$^2$ and Nuno F. Loureiro$^2$\\
$^{1}$ Department of Applied Mathematics, University of Leeds, Leeds, LS2 9JT, UK \\
$^{2}$ Plasma Science and Fusion Center, Massachusetts Institute of Technology, Cambridge, MA 02139, USA}
\begin{document}

\pagerange{\pageref{firstpage}--\pageref{lastpage}} \pubyear{2020}

\maketitle

\label{firstpage}

\begin{abstract}
It has been recently shown numerically 
that there exists an inverse transfer of magnetic energy in decaying, nonhelical, magnetically dominated, magnetohydrodynamic
turbulence in 3-dimensions (3D).  We suggest that magnetic reconnection is the underlying physical
mechanism responsible for this inverse transfer. In the two-dimensional (2D) case, the inverse
transfer is easily inferred to be due to smaller magnetic islands merging to form larger ones via
reconnection. We find that the scaling behaviour is similar between the 2D and the 3D cases, i.e.,
the magnetic energy evolves as $t^{-1}$, and the magnetic power spectrum follows a slope of $k^{-2}$.
We show that on normalizing time by the magnetic reconnection timescale, the evolution curves of
the magnetic field in systems with different Lundquist numbers collapse onto one another. Furthermore, transfer function 
plots show signatures of magnetic reconnection  driving the inverse transfer. We also discuss the
conserved quantities in the system and show that the behaviour of these quantities is similar
between the 2D and 3D simulations, thus making the case that the dynamics in 3D could be
approximately explained by what we understand in 2D. 
Lastly, we also conduct simulations where the magnetic field is subdominant to the flow. Here, too, we find an inverse transfer of magnetic energy in 3D. In these simulations, the magnetic energy evolves as $ t^{-1.4}$ and, interestingly, a dynamo effect is observed.
\end{abstract}


\section{Introduction}

Turbulent processes are of fundamental importance to a wide range of systems, from quantum fluids to
astrophysical plasmas \citep{SS2012, Biskamp2003}.  In a typical turbulent system, energy injected
at a certain scale \textit{direct} cascades down to smaller and smaller scales until it is
dissipated by  microphysical processes. On the other hand, an \textit{inverse} cascade, or inverse
transfer, involves energy being transferred from smaller to larger scales.  This can occur in both
forced or freely decaying turbulent systems \citep[e.g.,][]{Davidson2004}.  The
best known inverse cascading
system is two-dimensional (2D) hydrodynamic turbulence, where energy inverse cascades, while enstrophy
direct cascades \citep{Batch1969,Kraichnan1967}.  Indeed, the 2D hydrodynamic inverse cascade is widely considered one of the most
important results in turbulence \citep{Frisch1995, FS2006} since Kolmogorov's 1941 work.  Both
energy and enstrophy are inviscid invariants in 2D hydrodynamics. Here, the existence of more than
one ideally conserved quadratic quantity in the system can lead to an inverse cascade
\citep{Naz2011}.  The 3D system mimics 2D-like inverse transfer when there is anisotropy due to
strong rotation or the presence of a strong magnetic field
\citep{YakhotPelz87,Baggaleyetal2014,Pouquetetal2019}.  \citet{Biferaleetal2012}
demonstrated that even in the case of 3D isotropic and homogeneous hydrodynamic turbulence, there can
be an inverse cascade when parity (mirror symmetry) of the flow is broken. 

Similarly, in 3D magnetohydrodynamics (MHD),
it is well known that even in isotropic and homogenous
decaying turbulence, inverse cascade occurs due to the presence of non-zero net magnetic helicity which
breaks the parity in the system \citep{PFL76,CHB2001}.  Magnetic helicity is a
well-conserved quantity in the limit of large magnetic Reynolds number ($\Rm$).  Thus it is possible
to have an inverse transfer in decaying turbulence in 3D MHD as long as it has helical magnetic
fields \citep{CHB2001}.  However, recent simulations~ \citep{axel2015,zrake2014,bereraLinkmann2014,reppinbanerjee2017,munietal2020} have shown that there exists an inverse transfer of
magnetic energy in 3D MHD decaying turbulence, even in the absence of magnetic helicity.  

In this paper we investigate the underlying cause of such a 3D nonhelical inverse transfer.  We find
that there are similarities between the 2D and 3D cases. The 2D inverse transfer has been previously
well-studied and the ideal conserved quantities have been identified such as the total energy and
vector-potential squared \citep{FM76, pouquet_1978, BW1989}.  However, earlier 2D studies used
Kolmogorov-type arguments to obtain scaling solutions for the decaying field \citep{Biskamp2003}.
These arguments do not shine light upon the underlying physical processes responsible for the
inverse transfer.  In recent work by \citet{Munietal2019}, a simple model based on merging
magnetic islands provides a physical picture for the inverse transfer in the 2D system, and finds that
the relevant timescale is that dictated by magnetic reconnection, which underlies such mergers. Here, we propose that magnetic reconnection is responsible for the 3D nonhelical inverse transfer as well. 
Using direct numerical simulations, we study 3D, nonhelical, decaying MHD turbulence and build
connections to the 2D case.  We present evidence of similarities between 2D and 3D systems, and suggest magnetically dominated 3D systems display a
2D-like behavior.

We believe these findings are pertinent to several cosmological and astrophysical contexts. 
This reconnection-based understanding of the nonhelical inverse transfer, if true, affects the timescales of magnetic field evolution in the early universe \citep{banerjeeJedamzik2004, sethikandu2005, kandu2016}.  
Occurrence of reconnection in magnetically dominated decaying turbulence can be relevant to the understanding of high energy phenomena 
such as gamma-ray bursts and Crab nebula flares \citep{asanoTerasawa2015, zrake2016, blandfordetal2017}. 
Furthermore, such decaying turbulence has been studied in
the context of star-formation in molecular clouds \citep{maclowetal1998,gaoetal2015}, and is
relevant to the seeding of magnetic fields in protogalaxies from supernovae ejecta \citep{becketal2013},
and also in the case of galaxy-clusters after a merger event \citep{sss2006, sur2019}. 
\cite{munietal2020} present a discussion on the significance of inverse transfer in obtaining seed magnetic fields required for galactic dynamos (see their Appendix A). 

\section{Numerical setup}
\label{numset}
\subsection{The model}
We use the \textsc{Pencil Code}\footnote{DOI:10.5281/zenodo.2315093, github.com/pencil-code}  
to simulate decaying MHD turbulence in both 2D and 3D. 
We solve the MHD equations given by 
\EQ {D\ln\rho\over Dt}=-\nab\cdot\uu, 
\label{dlnrhoS}\EN
\EQ {D\uu\over D
t}= - c_{\rm s}^2 \nab \ln\rho + \frac{\JJ\times\BB}{\rho} + \frac{\bm{F}_{\rm visc}}{\rho}, 
\label{dUUS} \EN 
\EQ
{\partial \AAA\over \partial t}=\uu\times\BB-\eta\mu_0\JJ, 
\label{dAdt}
\EN
on a Cartesian $N^2$ or $N^3$ grid, with periodic boundary conditions, where $N$ is the number
of grid points in any given direction. 
The operator $D/Dt=\partial/\partial t + \uu\cdot \nab$ is the advective derivative, with $\uu$ the
fluid velocity field.  We solve the uncurled version of the induction equation, in terms of the
vector potential, $\AAA$, related to the magnetic field by $\BB=\nab \times \AAA$.  We adopt the Weyl
gauge $\Phi=0$, where $\Phi$ denotes the scalar potential.  The current density is $\JJ=\nabla
\times \BB/\mu_0$, with $\mu_0$, the vacuum permeability. 
The viscous force is $\FF_{\rm visc} = \nab\cdot2\nu\rho\SSSS$,
where $\nu$ is the kinematic viscosity,
and $\SSSS$ is the traceless rate of strain tensor with components
${\SSSS}_{ij}=\frac{1}{2}(u_{i,j}+u_{j,i})-\frac{1}{3}\delta_{ij}\nab\cdot\uu$ (commas denote partial derivatives).
Finally, $\eta$ is the magnetic diffusivity. 
In the 2D runs, we solve a 2D version of  equations (\ref{dlnrhoS} - \ref{dAdt}) obtained by setting $\partial_z=0$ and eliminating  vector components in the $z$ direction. Other than compressibility effects (which are minor in our simulations), this 2D version of the equations is identical to the 2D version of the reduced-MHD equations~\cite{kadomtsev_1974,strauss_1976,alex_2009}.
The code uses dimensionless quantities by
measuring length in units of the domain size $L$, speed in units of the isothermal sound speed $c_{\rm s}$, density in units of the initial value $\rho_0$ and magnetic field in units of $(\mu_0\rho_0
c_{\rm s}^2)^{1/2}$. We choose $L=2\pi$, and  $c_s=\rho_0=\mu_0=1$.

\subsection{Initial conditions and parameters}
\label{sec:ic_and_param}
The initial magnetic field is generated in the wavenumber space with a certain spectrum and random
phases, similar to the method in \citet{axel2015}. The magnetic power spectrum is $k^4$ \citep{axel2015} for $k<k_0$,
and is exponentially cutoff beyond $k_0$. Such a spectrum is obtained from the vector potential in
Fourier space, $\hat{A_j}(\kk)$, whose three components $j$ are given by, 
\EQ
\hat{A_j}(\kk)= A_0 (k^2/k_0^2)^{n/4-1/2} \exp{(-k^2/k_0^2)} \exp{(i\phi(\kk))}
\label{initcond}
\EN
where exponent $n=2$, $\phi(\kk)$ are random phases and $A_0$ is the amplitude.

We define the Lundquist number in our simulations as $S=V_A (2\pi/k_p)/\eta $, where $V_A$ is the Alfv\'en velocity and
$k_p$ is the wavenumber at which the magnetic power spectrum peaks. 
In our main runs, the initial Alfv\'en
velocity is $V_A=0.2$ (which implies that compressibility effects are weak and can be ignored in the analysis of the dynamics) and the initial velocity field is zero, analysed in sections \ref{2Dsims}-\ref{tfrfunc}.  We have also carried out runs with non-zero initial velocity field; these are reported in section \ref{nonzerov}.
In all runs, initial $k_p\approx25$.  We have run simulations across a range of Lundquist
numbers as allowed by the resolution limit of $2048^2$ in 2D and $1024^3$ in 3D. 
The magnetic Prandtl number in all 
our simulations is $1$. A list of all runs reported in this paper is shown in Table~\ref{xxx}.

\begin{table}
\setlength{\tabcolsep}{8.0pt}
\caption{A summary of all runs and their respective parameters (in dimensionless units). $u_{rms0}$ 
and $B_{rms0}$ are the initial root-mean-squared values of the flow and the magnetic field respectively. In all runs the initial $k_p\approx25$.}
 \begin{tabular}{|c|c|c|c|c|c|}
\hline
\hline
Run & {\small Resolution} & $\eta\times10^4$ & $u_{rms0}$ & $B_{rms0}$ & $S$ \\ 
\hline
A2D & $2048^2$  & 0.5  &   0.0  & 0.2  &  1000   \\
B2D & $1024^2$  & 1.0  &   0.0  & 0.2  &  500   \\
C2D & $1024^2$  & 2.0  &   0.0  & 0.2  &  250   \\
F2D & $1024^2$  & 1.0  &   0.2 & 0.02 &  50   \\
A3D & $1024^3$  & 0.5  &   0.0  & 0.2  &  1000   \\
B3D & $1024^3$  & 1.0  &   0.0  & 0.2  &  500   \\
C3D & $512^3$   & 2.0  &   0.0  & 0.2  &  250   \\
D3D & $512^3$   & 4.0  &   0.0  & 0.2  &  125   \\
E3D & $512^3$   & 8.0  &   0.0  & 0.2  &  50   \\
F3D & $1024^3$  & 0.25 &   0.2 & 0.02 &  200    \\
\hline
\hline
\label{xxx}
\end{tabular}
\end{table}

\section{Results}
\label{results}

\subsection{Decaying turbulent magnetic fields in 2D}
\label{2Dsims}

\label{result}
\begin{figure}
\centering
\includegraphics[width=0.35\textwidth, height=0.17\textheight]{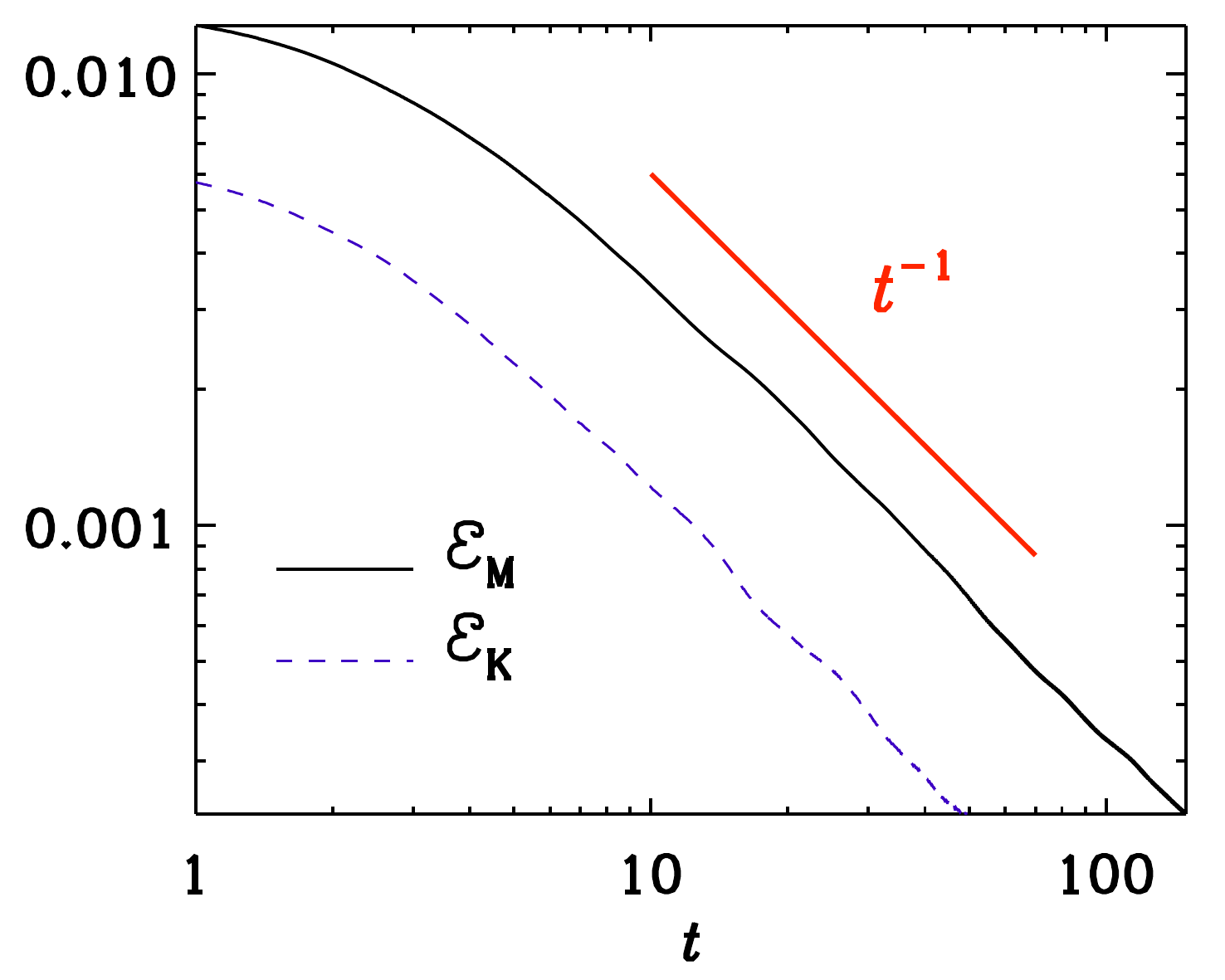}
\includegraphics[width=0.35\textwidth, height=0.17\textheight]{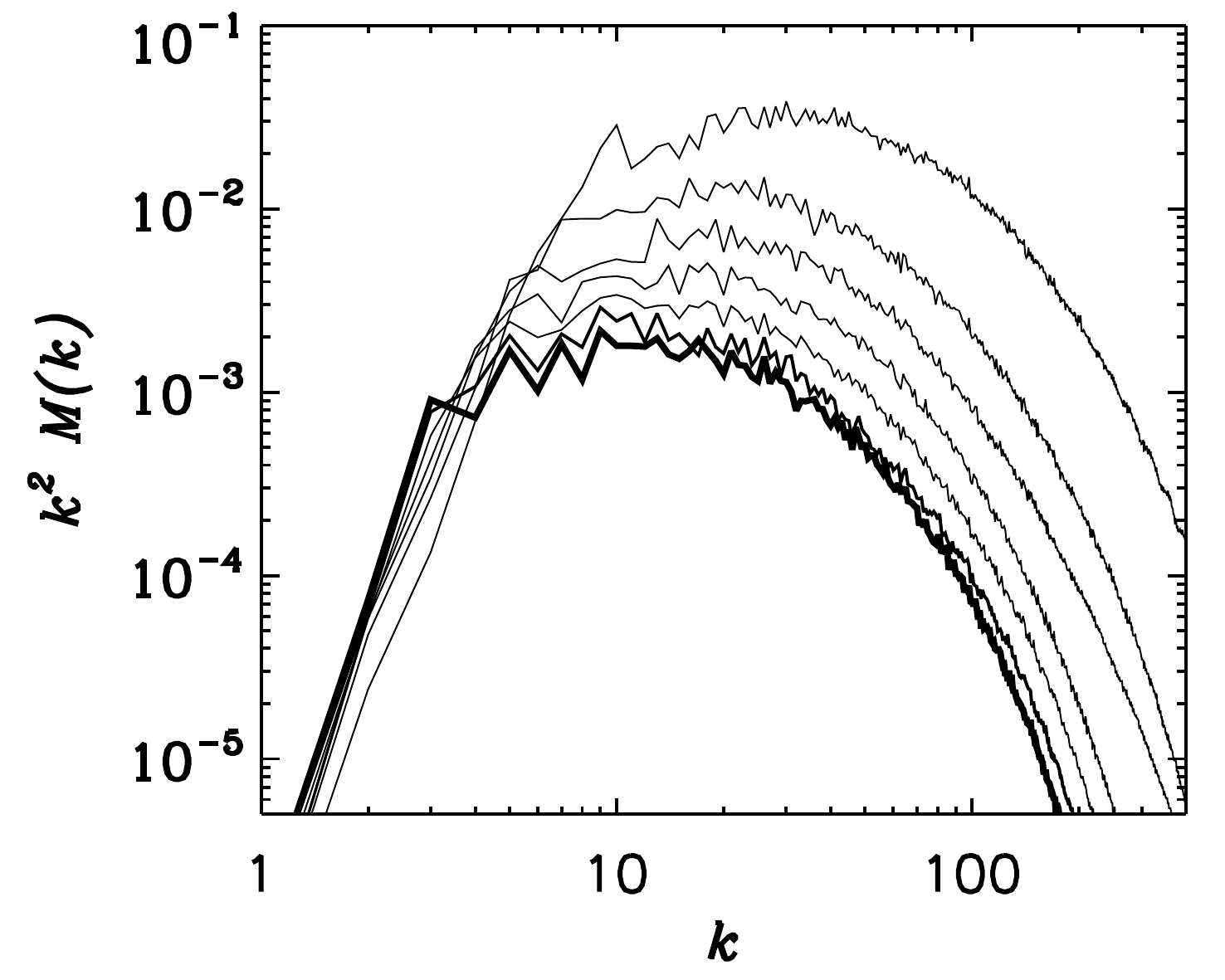}
\caption{
Top panel: Evolution of magnetic energy (solid black) and kinetic energy (dashed blue) in a 2D simulation, A2D, with $S=1000$ and resolution of $2048^2$.
Bottom panel: Compensated magnetic power spectra $k^2 M(k,t)$ are plotted at regular intervals of $\Delta t=10$ with a thick final curve at $t=70$.
}
\label{2Devol}
\end{figure}

\begin{figure*}
\includegraphics[width=0.33\textwidth, height=0.21\textheight]{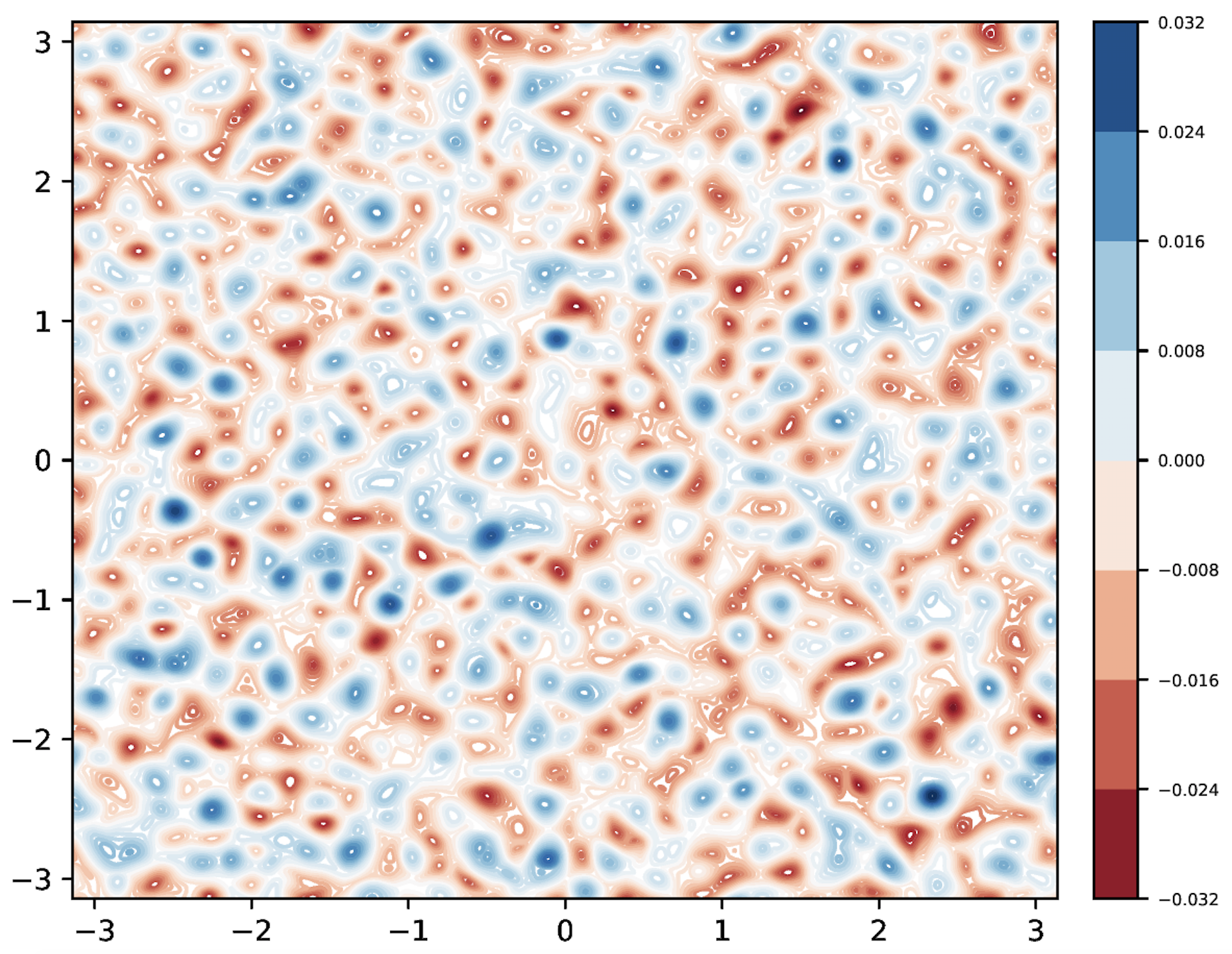}~~
\includegraphics[width=0.33\textwidth, height=0.21\textheight]{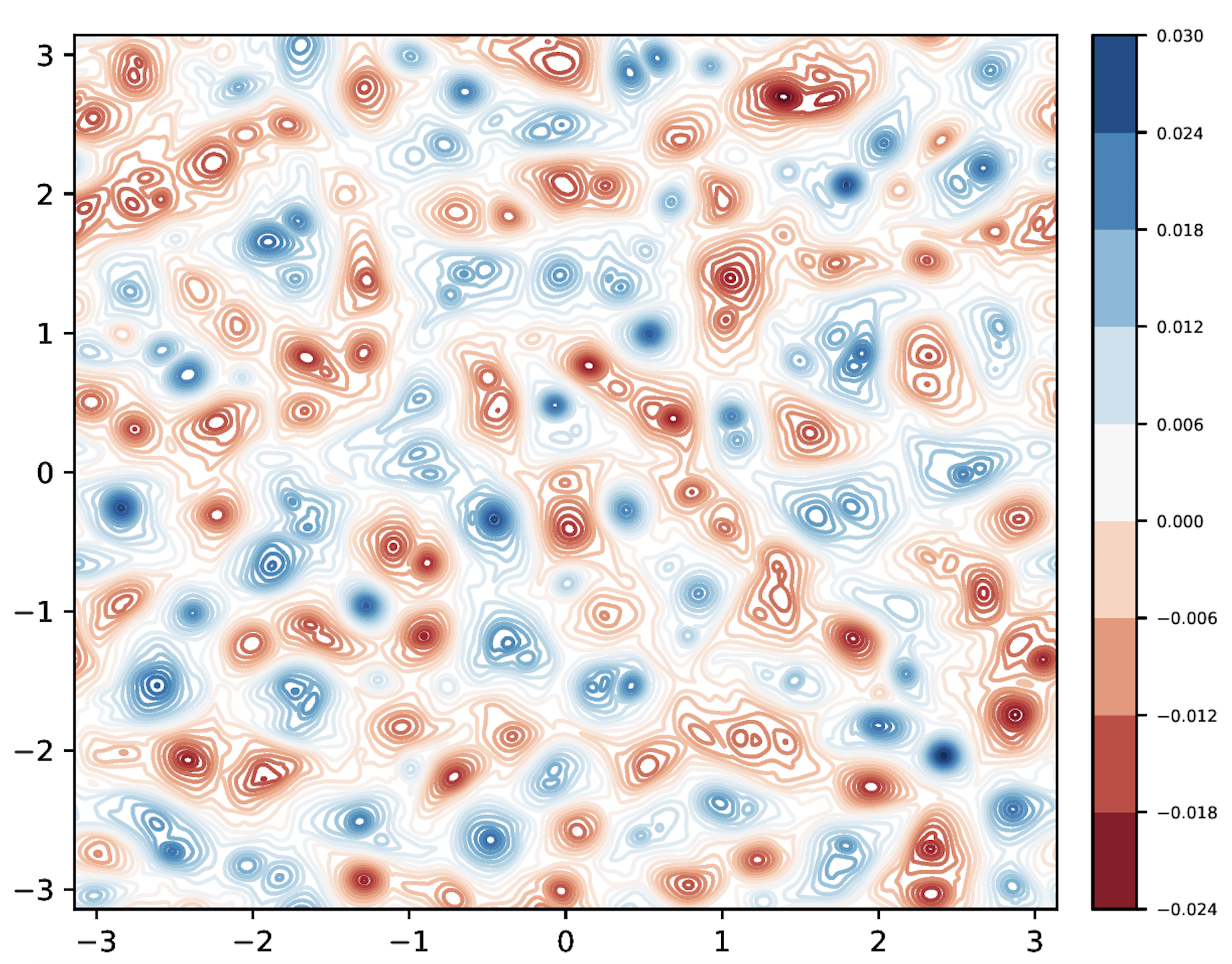}~~
\includegraphics[width=0.33\textwidth, height=0.21\textheight]{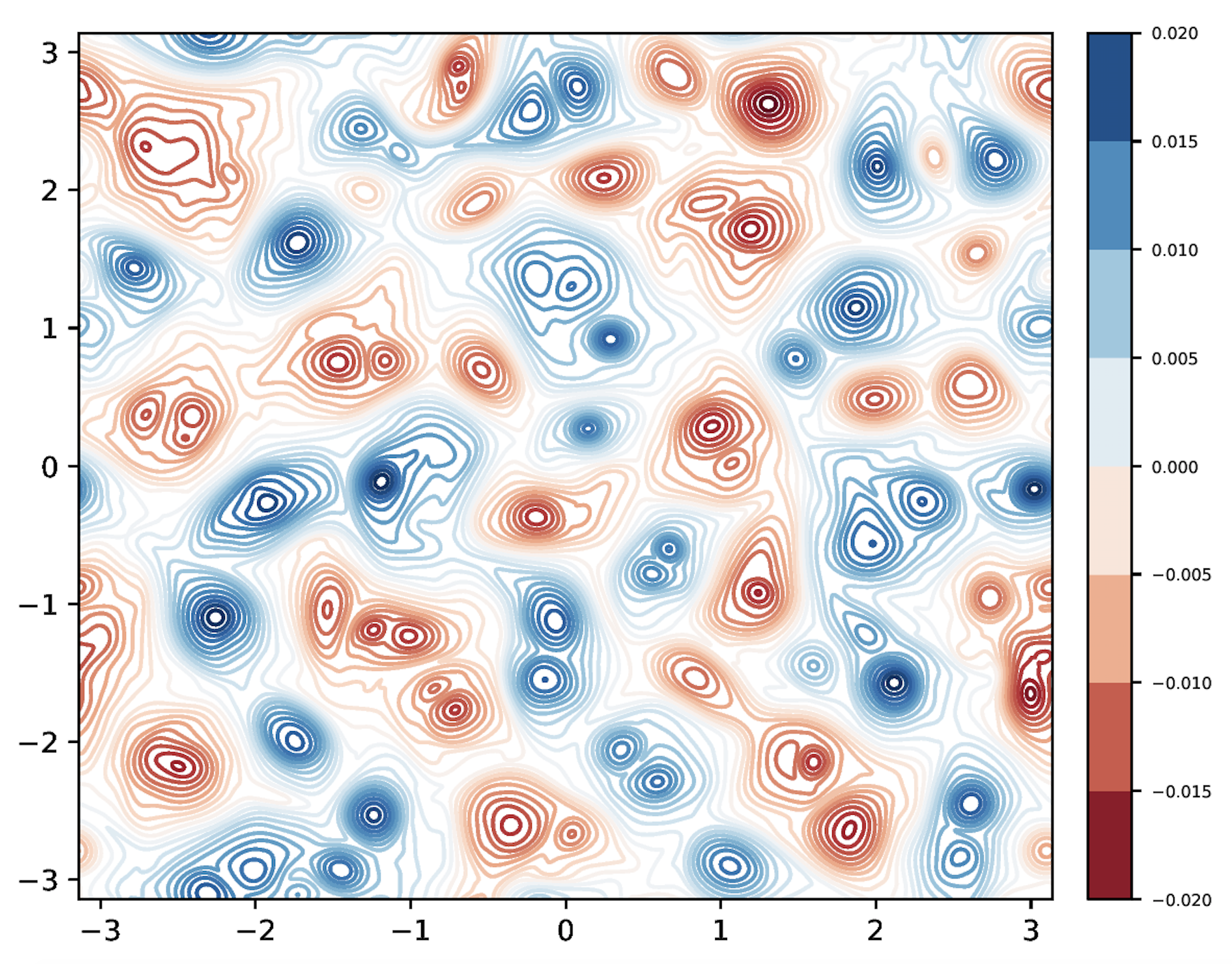}
\caption{
Evolution of the vector potential ($A_z$) in the 2D simulation A2D, with $S=1000$. 
The times plotted are $t=1$, $t=15$ and $t=45$, from left to right. }
\label{contevol}
\end{figure*}

\begin{figure*}
\includegraphics[width=0.33\textwidth, height=0.21\textheight]{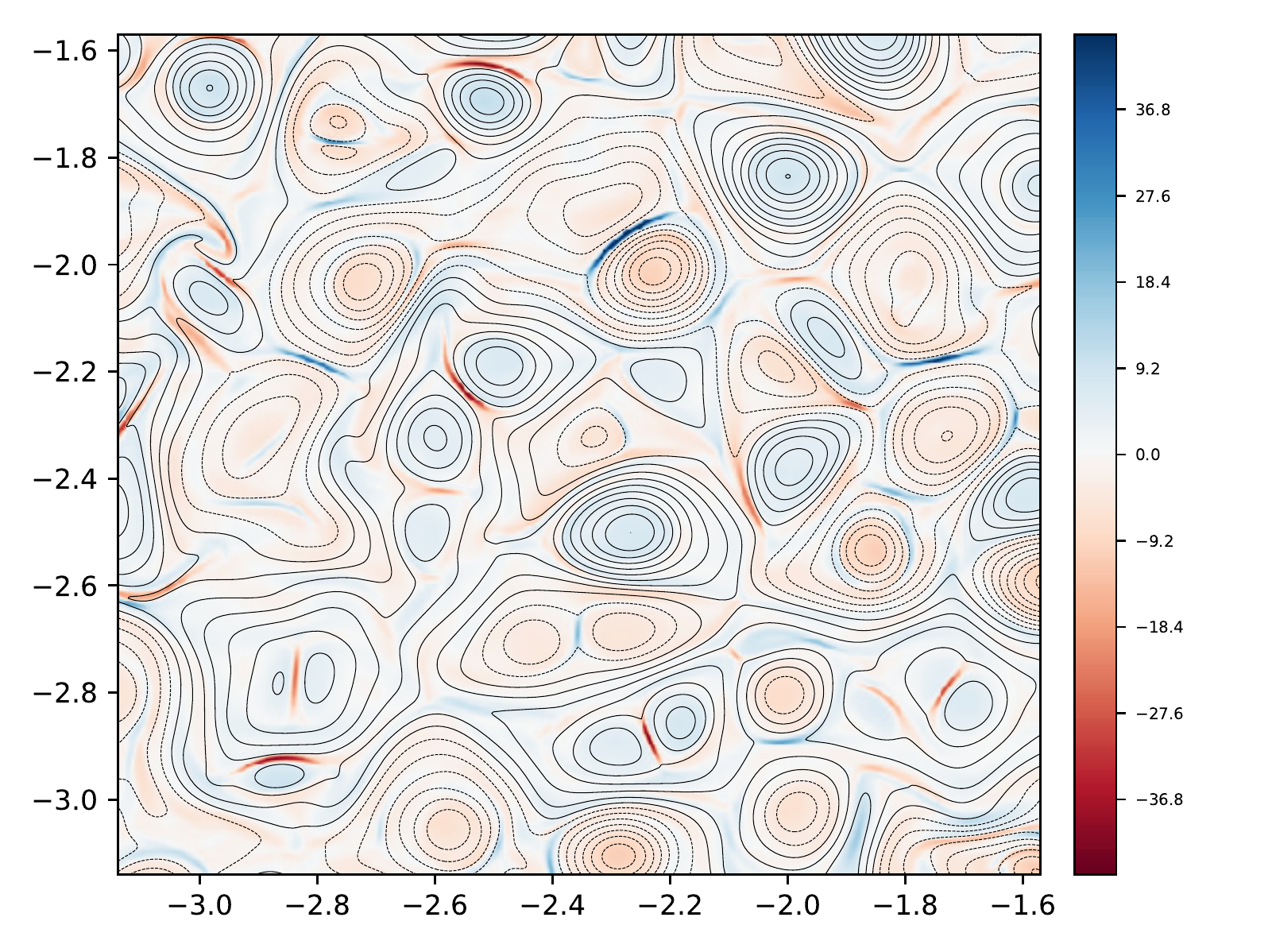}~~
\includegraphics[width=0.33\textwidth, height=0.21\textheight]{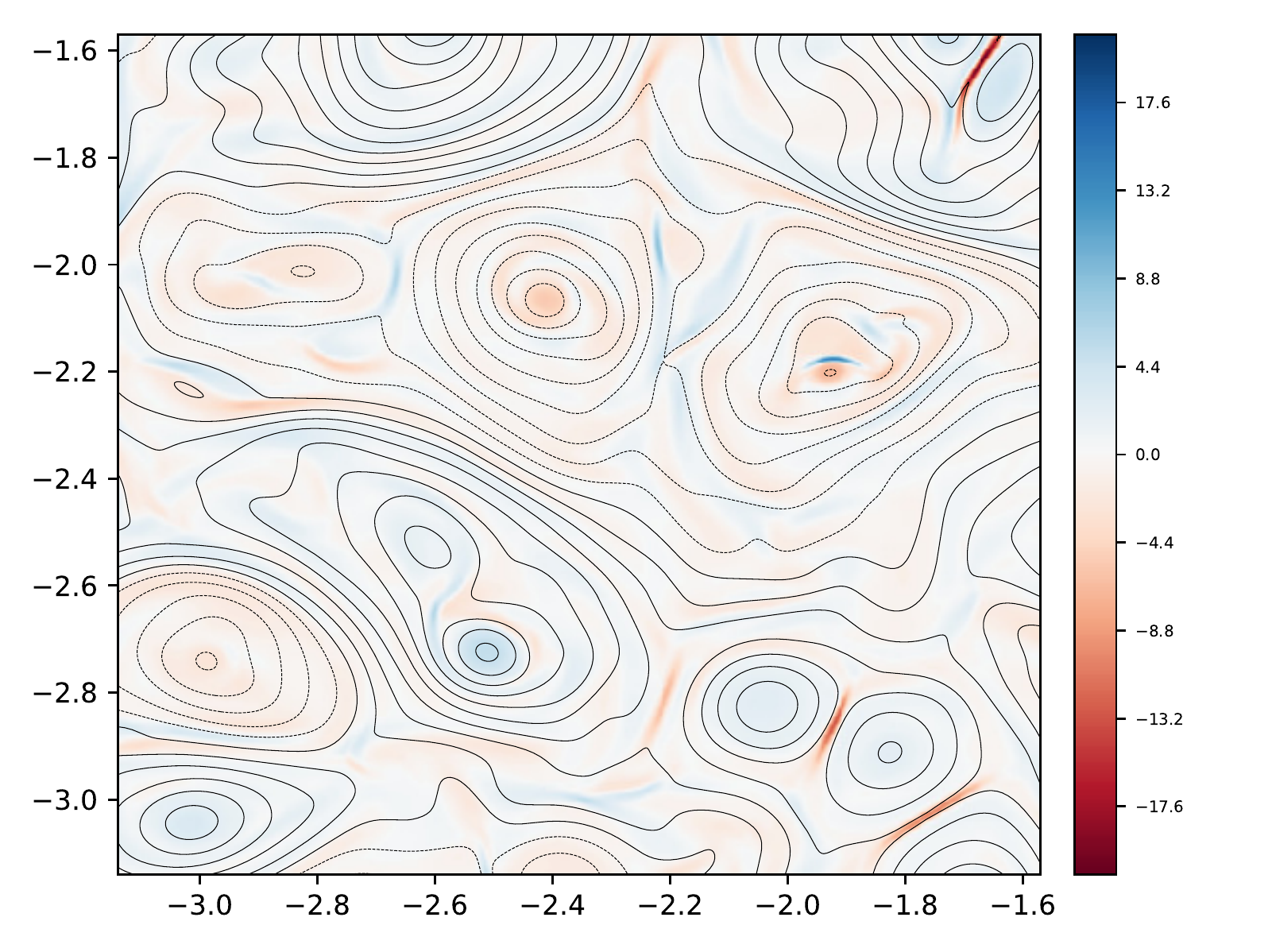}~~
\includegraphics[width=0.33\textwidth, height=0.21\textheight]{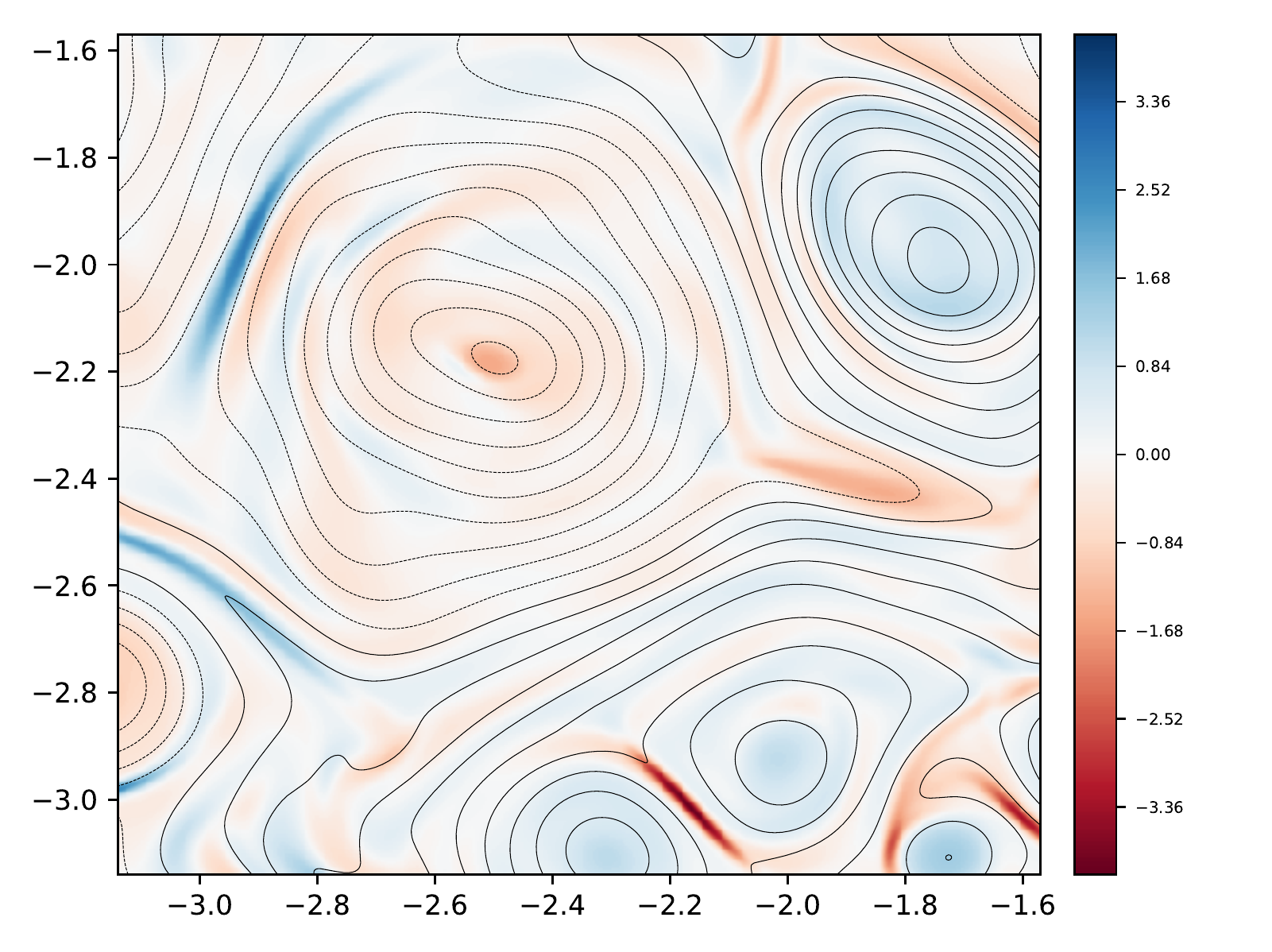}
\caption{Evolution of the current density ($J_z$) in $1/8$ of the domain from the 2D simulation A2D, at times $t=1$, $t=15$ and $t=45$, from left to right. The overlaid lines are contours of $A_z$.}
\label{contjevol}
\end{figure*}

We first present our study of 2D simulations of decaying MHD turbulence (simulations A2D, B2D and C2D in Table~\ref{xxx}). 
In the top panel of \Fig{2Devol}, the evolution of magnetic energy is
shown in a log-log plot. It decays in time as a power law, with an exponent
close to $-1$ at late times.  This result matches with that obtained in the 2D
simulations performed by \citet{Munietal2019}, which focused on an initial condition consisting of an ordered array of current filaments (or, equivalently, magnetic islands)
with alternating polarities. 
Upon introducing small perturbations into that system, the current filaments
move out of the initial (unstable) equilibrium. The subsequent evolution of the system is then primarily dictated by the coalescence, via magnetic reconnection, of filaments with equal polarity.
Mergers of island pairs lead to larger islands, albeit at
the cost of magnetic energy dissipation. 
Successive mergers lead to progressively larger structures,
resulting in an inverse transfer of magnetic energy.  This occurs hierarchically in a self-similar
manner, giving rise to power-law-in-time behaviour.  
Similarly, even in the case of a random initial condition such as employed here, we observe an
inverse transfer as the system evolves in time.  This is quite evident in the time progression of the magnetic power spectrum, shown in the bottom panel of \Fig{2Devol}. 
The initial spectrum (random field peaked at $k_p\sim25$) is seen to shift from large wavenumbers to smaller ones, depicting an inverse transfer. 
The spectra are compensated by $k^2$ to reveal a range where they flatten; the same power law is observed by~\citep{Munietal2019}, who attribute it to the dominance of sharp current sheets. 
(i.e., a Burgers' spectrum~\citep{burgers_1948}).

As in~\citet{Munietal2019}, the growth of magnetic energy at large scales that we find in our 2D simulations is due to magnetic reconnection.  This can be seen explicitly and clearly from a sequence (a movie) of time evolving
contour plots of $A_z$ (see supplementary material), or from the corresponding stills at specific moments of time shown in \Fig{contevol}.
Current sheets --- sharply localized enhancements of current density in \Fig{contjevol} ---  are seen to form at the
interface of any pair of interacting islands, leading to their reconnection and resulting in
larger islands.  
The magnetic islands can be seen to grow progressively larger in time due to
island mergers. 
 
\begin{figure}
\centering
\includegraphics[width=0.35\textwidth, height=0.17\textheight]{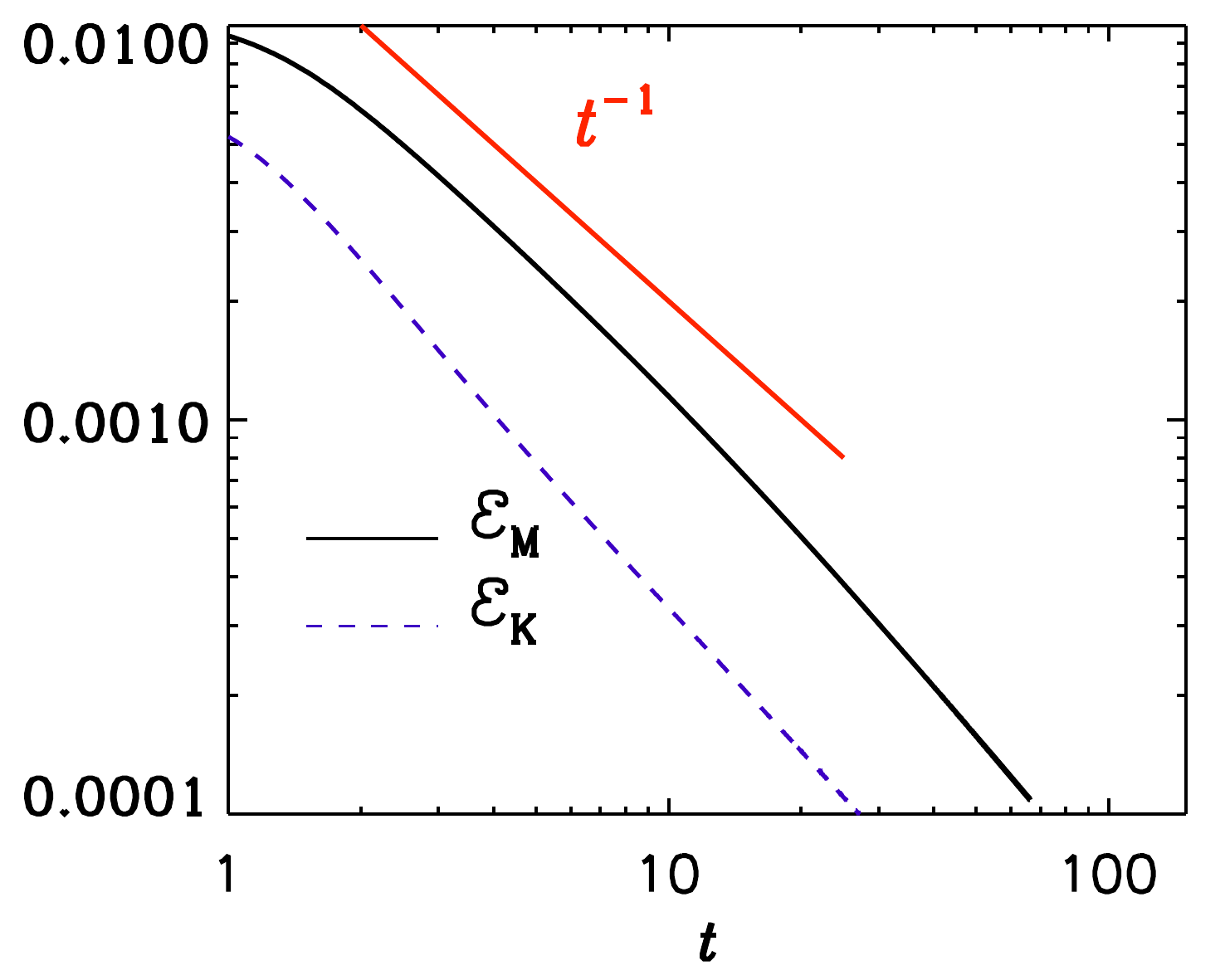}
\includegraphics[width=0.35\textwidth, height=0.17\textheight]{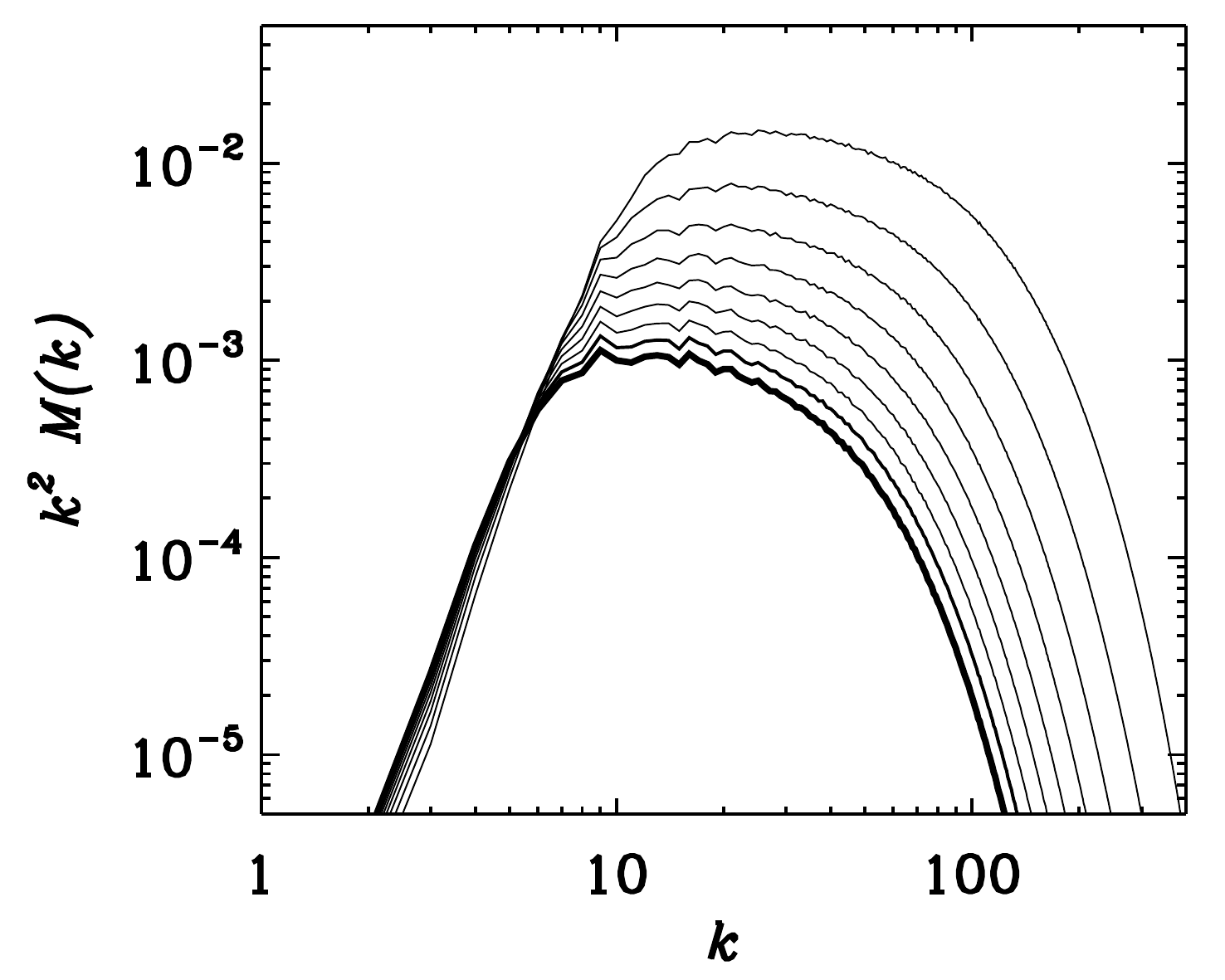}
\caption{
Top panel: evolution of magnetic energy (solid black) and kinetic energy (dashed blue) in the 3D simulation A3D, with
$S=1000$ and resolution of $1024^3$. The bottom panel shows compensated magnetic power spectra $k^2 M(k,t)$ for the same run, plotted at regular intervals of $\Delta t=5$, with a thick final curve at $t=50$.}
\label{3Devol}
\end{figure}

\begin{figure*}
\includegraphics[width=0.33\textwidth, height=0.21\textheight]{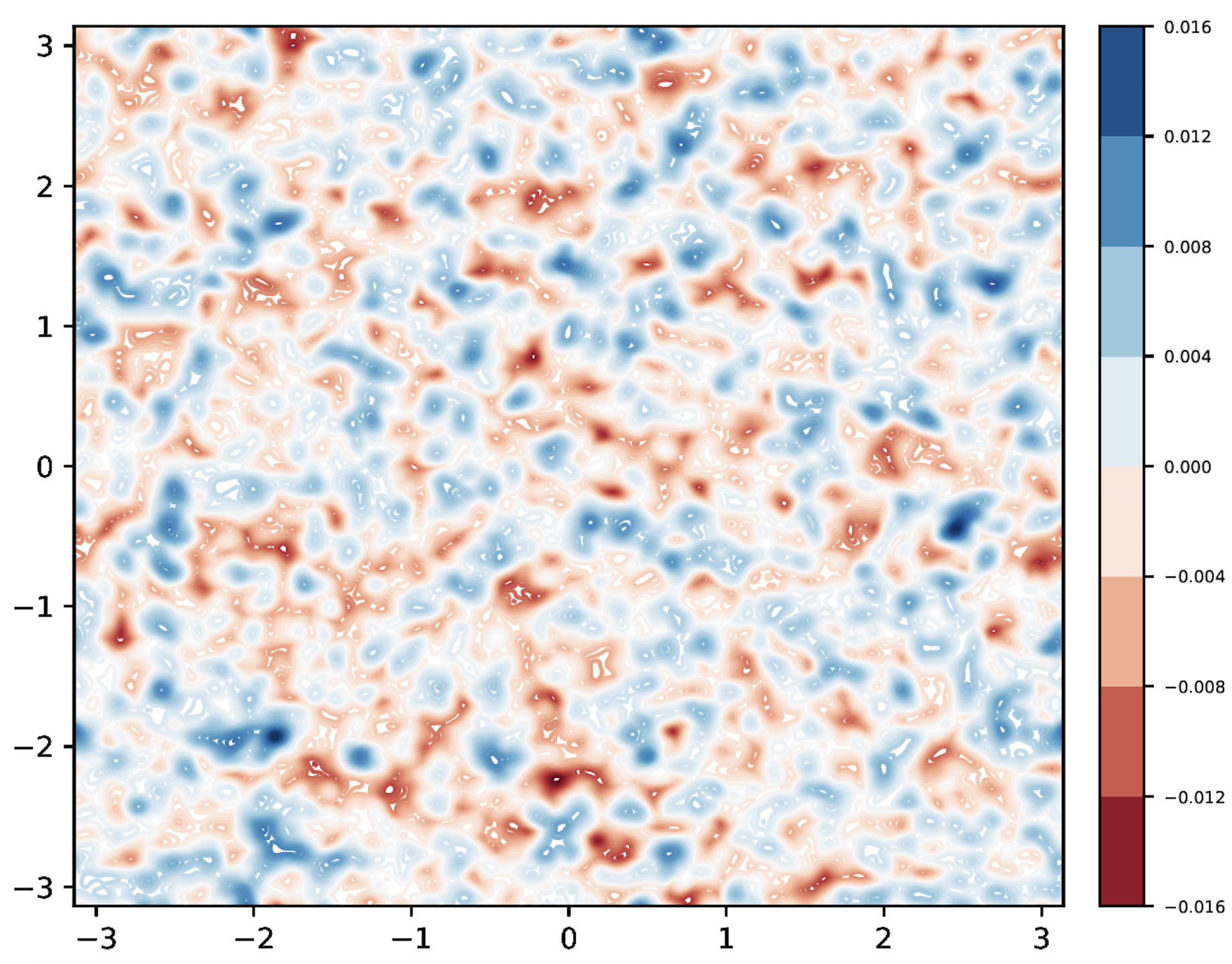}~~
\includegraphics[width=0.33\textwidth, height=0.21\textheight]{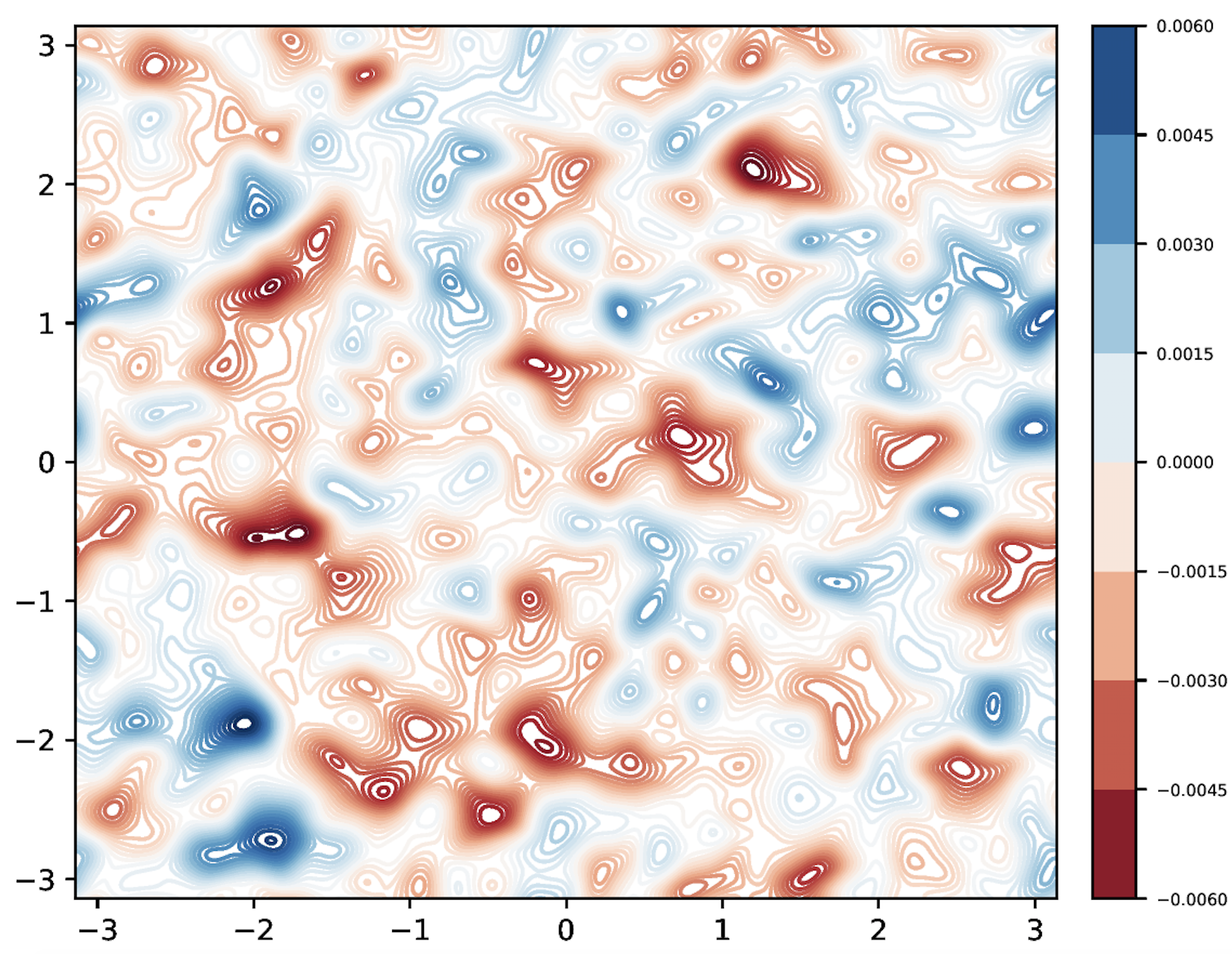}~~
\includegraphics[width=0.33\textwidth, height=0.21\textheight]{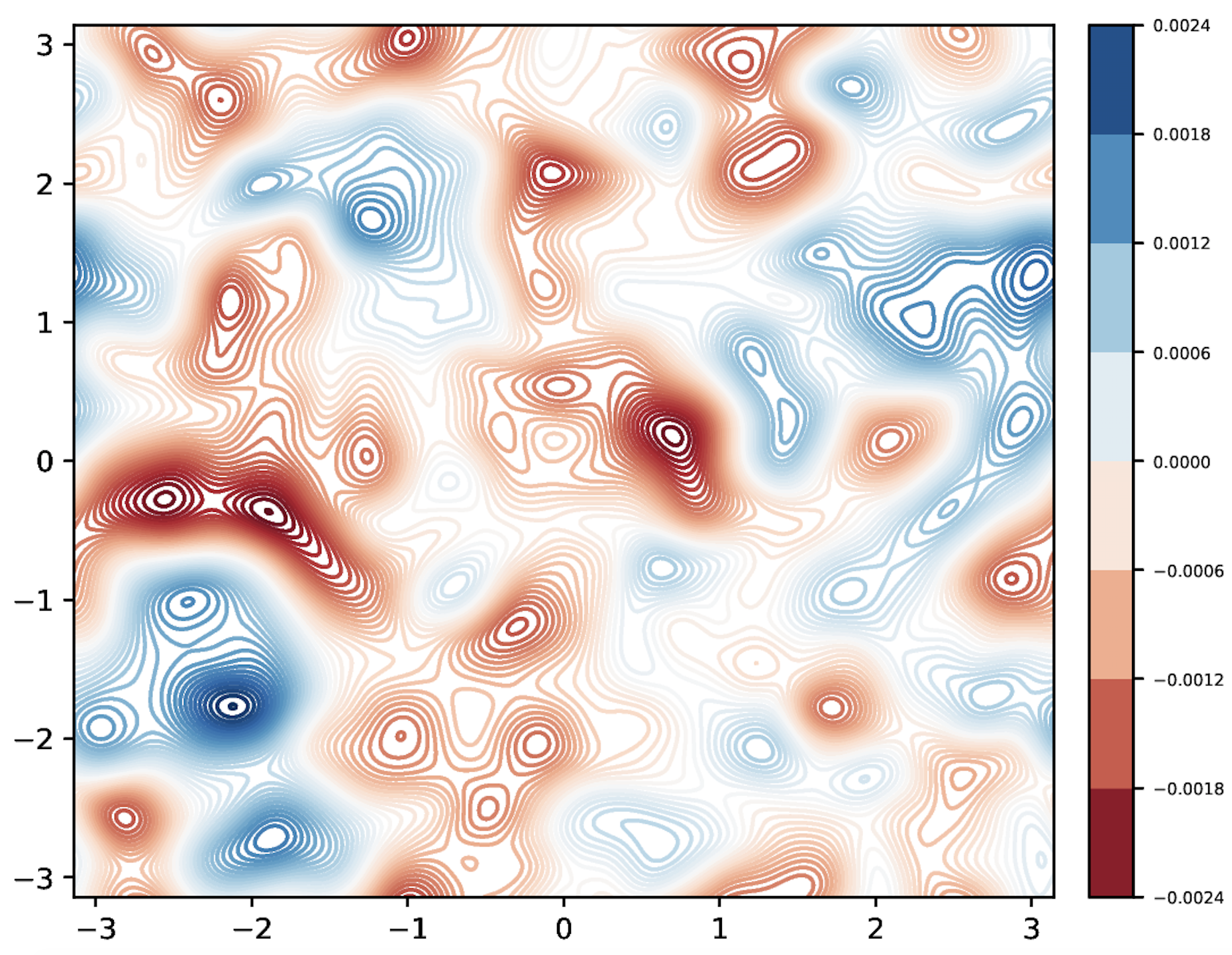}
\caption{
Contour plots of the $z$-component of the vector potential ($A_z$), in an arbitrary 2D slice (in the $x-y$ plane) from the 3D simulation C3D. The times plotted are 
$t=2$, $t=15$ and $t=50$, from left to right.}
\label{contevol3d}
\end{figure*}

\begin{figure*}
\includegraphics[width=0.33\textwidth, height=0.24\textheight]{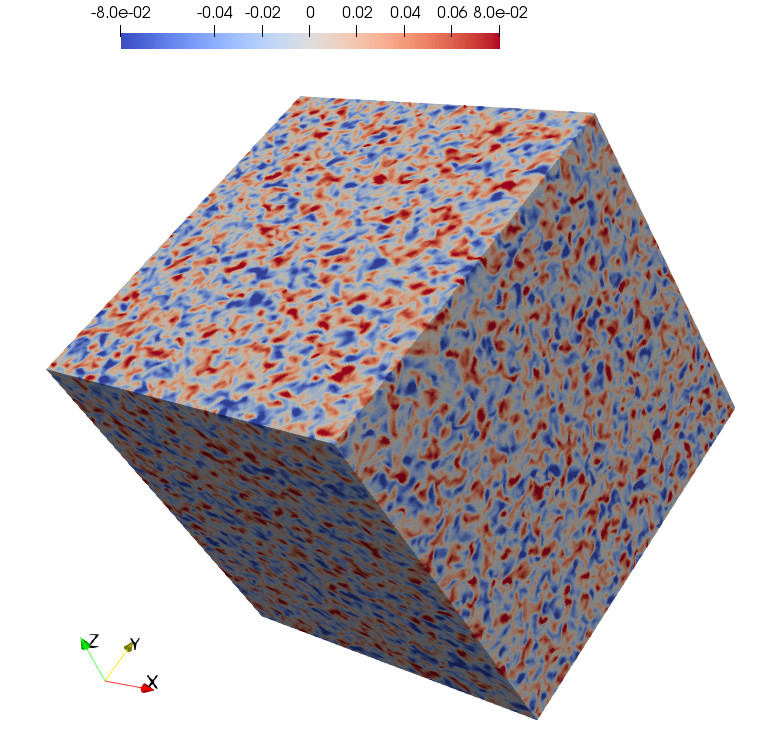}~~
\includegraphics[width=0.33\textwidth, height=0.24\textheight]{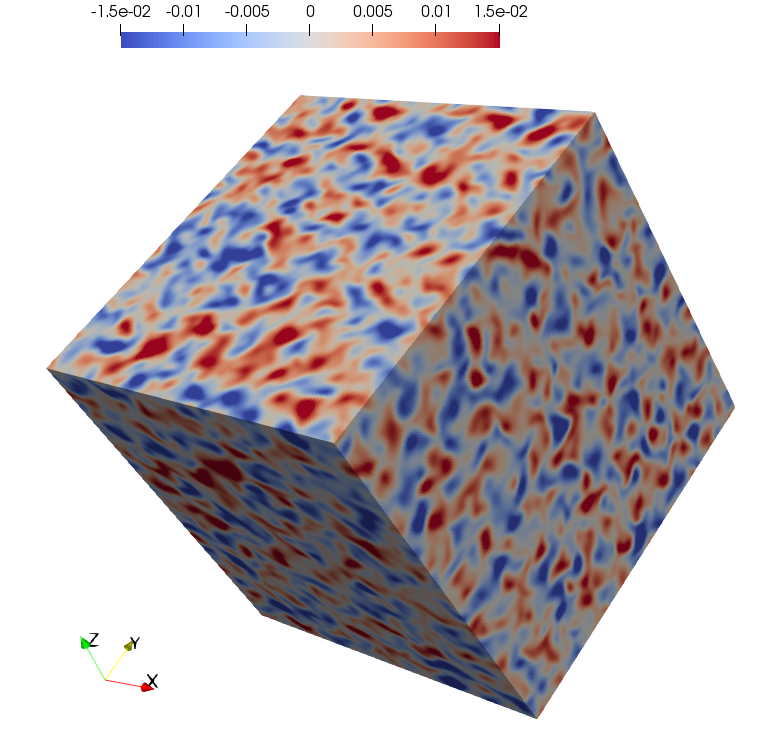}~~
\includegraphics[width=0.33\textwidth, height=0.24\textheight]{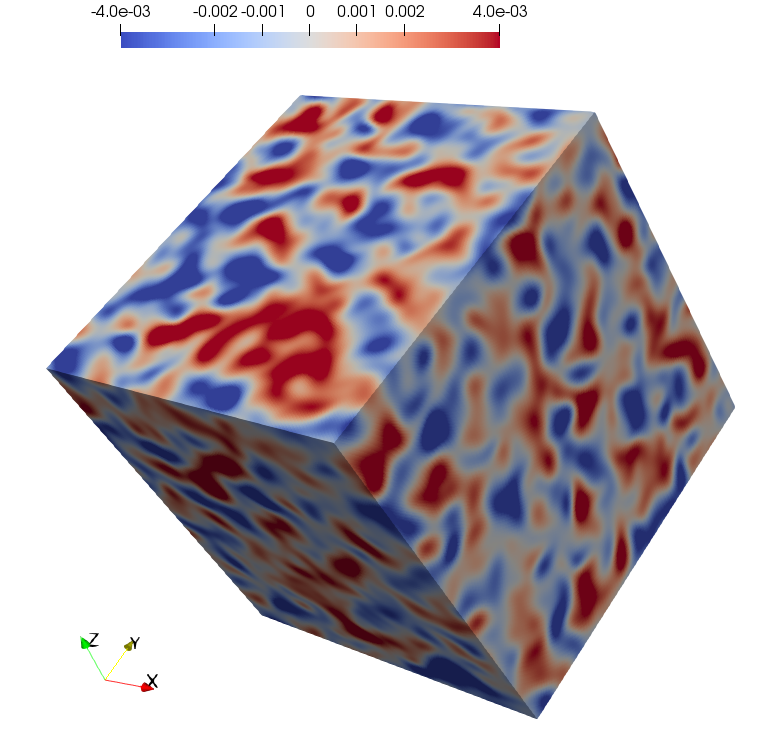}
\caption{
Evolution of a component of the magnetic field, $B_x$, shown on the 3D domain from the 3D simulation, C3D,  at times $t=2$, $t=15$ and $t=50$ from left to right.}
\label{boxevol3d}
\end{figure*}

A complementary way to understand inverse transfer in this 2D system is to consider the conserved quantities in the system.  For the ideal 2D MHD equations (in the absence of dissipation), these are the total energy, $\EEM+\EEK=\bra{\BB^2}/2 + \rho\bra{\uu^2}/2$ (given weak compressibility), and vector-potential squared, $P=\bra{\AAA^2}$, (where $\bra{}$ represents integral over the domain) \citep[e.g.,][]{Biskamp2003}. In the following we show that in our non-ideal system where the kinetic energy is subdominant, by considering the evolution of $\EEM$ and $P$, we can deduce that decaying 2D MHD turbulence displays inverse transfer of energy. The evolution equations for magnetic energy and vector-potential squared in a closed domain (given periodic boundaries in the DNS) in the non-ideal case are given by,  
\EQA
\label{cons1}
\iint dS~  \frac{\partial}{\partial t} \left(\frac{\BB^2}{2}\right) &=& -\iint dS~ \uu \cdot (\JJ \times \BB) + \eta \JJ^2, \\ 
\iint dS~ \frac{D}{Dt} \left(\frac{A_z^2}{2} \right) &=& -\iint dS~ \eta \BB^2. 
\label{cons2}
\ENA
In the 2D limit that we consider here, only one component of the vector potential is needed, 
i.e. $\BB=\nab\times A_z\zzz$, where the $\zzz$ is the unit vector orthogonal to the 2D plane. Similarly, only one component of the current density survives, $J_z=\partial_x B_y -\partial_y B_x$.  

From \Eqs{cons1}{cons2}, it is possible to deduce the following implications for a freely evolving turbulent system. 
While the evolution of vector potential squared is governed by only a decay term on the RHS of \Eq{cons2}, 
the equation for magnetic energy, \Eq{cons1}, also consists of a source term given by $\uu \cdot (\JJ \times \BB)$. 
Depending on the sign of this term, 
either the energy is being transferred from the magnetic field to the velocity field, or {\it vice versa}. 
Now, in these simulations, the velocity field is initially zero, and it is entirely 
driven by the magnetic field. We assume that the back-reaction from the generated flow on the field is negligible:
this is a reasonable assumption if the kinetic energy is subdominant, as is indeed the case in our system (see the top panel of \Fig{2Devol}).
Thus, as the system is allowed to evolve freely, the magnetic field loses its energy to either the velocity field or to resistive decay. 
Given that the system is turbulent, the magnetic field is expected to decay even as $\eta \to 0$
because the field can develop small enough scales (current sheets). As a result, $\eta \bra{\JJ^2}$ can remain finite  in that limit. 
However, as $\eta \to 0$, the term on the right hand side of \Eq{cons2}, $\eta \bra{\BB^2}$ (where $\bra{\BB^2}$ is essentially independent of resistivity) will go to zero, 
thereby rendering volume-integrated vector-potential squared, $P$, to be nearly invariant. 
In short, in the limit of $\eta\to 0$ (or, equivalently, in the limit of very large $\Rm$ or $S$), the vector potential squared is better conserved than the magnetic energy, $\EEM=\bra{\BB^2}/2$. 

We can now use this conservation property to argue why such a freely decaying turbulent system can exhibit inverse transfer. 
In the Fourier domain we have $\hat{\BB}=i\kk\times \hat{A}\zzz$, where $\kk=k_x\xxx+k_y\yyy$. It follows that  
\EQ
\vert \hat{\BB} \vert^2 = k^2 \vert \hat{A} \vert^2.
\label{2DBArel}
\EN

Now, let us use the expressions for $\EEM$ and $P$ in the Fourier domain,
\EQA
\label{energy}
\EEM &=& \frac{1}{2}\int \vert \hat{\BB} \vert^2 ~~d^2\kk, \\
\label{vectorpotentialsquared}
P &=& \int \vert \hat{A} \vert^2 ~~ d^2\kk,
\ENA
and consider that most of the magnetic energy is concentrated in a single scale in the system; we shall call it the correlation scale, $k_{\rm corr}$. It then follows 
from \Eq{2DBArel} that 
\EQ
k_{\rm corr}\sim\sqrt{\EEM/P}.
\label{kcorr} 
\EN

Since this is an unforced stochastic system, the magnetic energy, $\EEM$, will decay. 
Given that $P$ is better conserved than $\EEM$, $P$ remains nearly constant as $\EEM$ decreases; thus, the wavenumber $k_{\rm corr}$ is expected to decrease. This implies a shift of 
the correlation scale in the system to larger and larger scales --- the spectral signature of an inverse transfer. Indeed, if we substitute the scaling $\EEM \propto t^{-1}$ 
into \Eq{kcorr}, and  consider $P$ to be constant in time, we obtain $k_{\rm corr}\propto t^{-1/2}$. This is consistent with what we find from our simulations when we trace $k_{\rm corr}$ as a function of time. 
Both of these scalings are predicted by the  reconnection-based hierarchical model of ~\citet{Munietal2019}, and are verified by the RMHD numerical simulations carried out in that paper (note that the hierarchical model itself is based on mass and magnetic flux ($A_z$) being conserved during island mergers through reconnection). 
Thus, we conclude that the implications from 2D conservation properties are consistent with the physical picture of island mergers via reconnection in 2D; together, they provide a solid explanation for the inverse transfer of magnetic energy in the 2D system. 

\subsection{Decaying turbulent nonhelical magnetic fields in 3D}
\label{3Dsims}

\begin{figure*}
\includegraphics[width=0.33\textwidth, height=0.21\textheight]{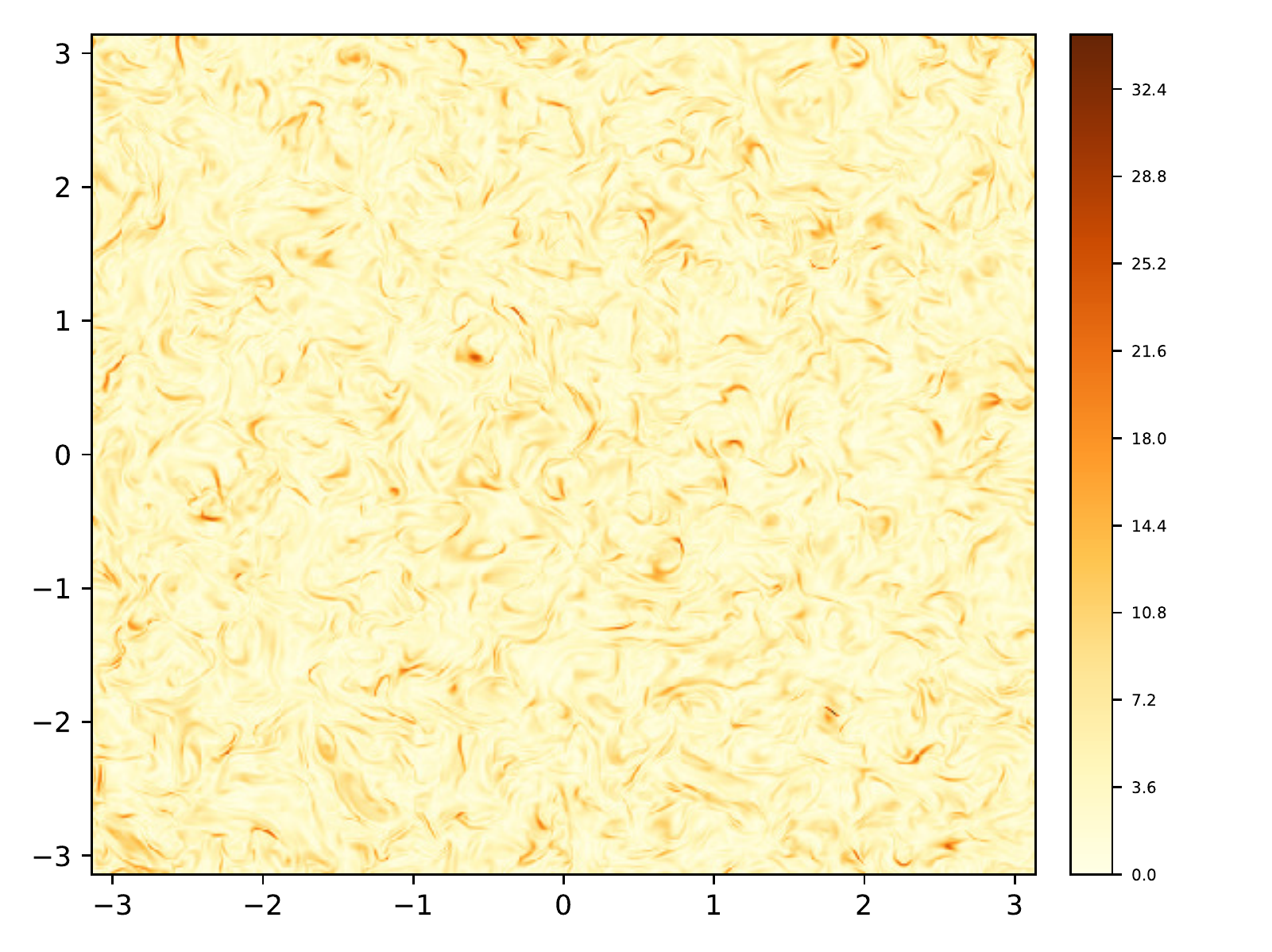}~~
\includegraphics[width=0.33\textwidth, height=0.21\textheight]{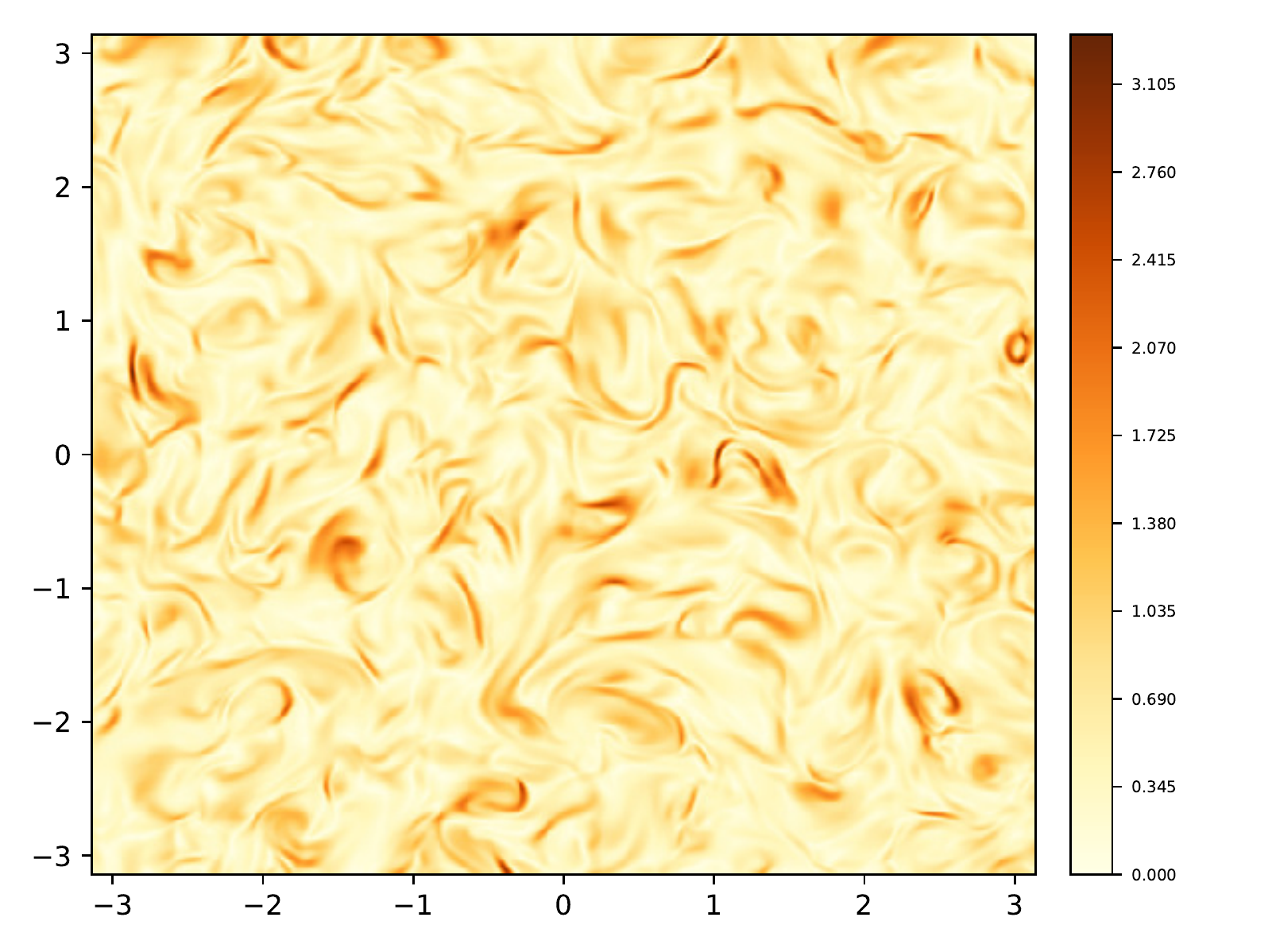}~~
\includegraphics[width=0.33\textwidth, height=0.21\textheight]{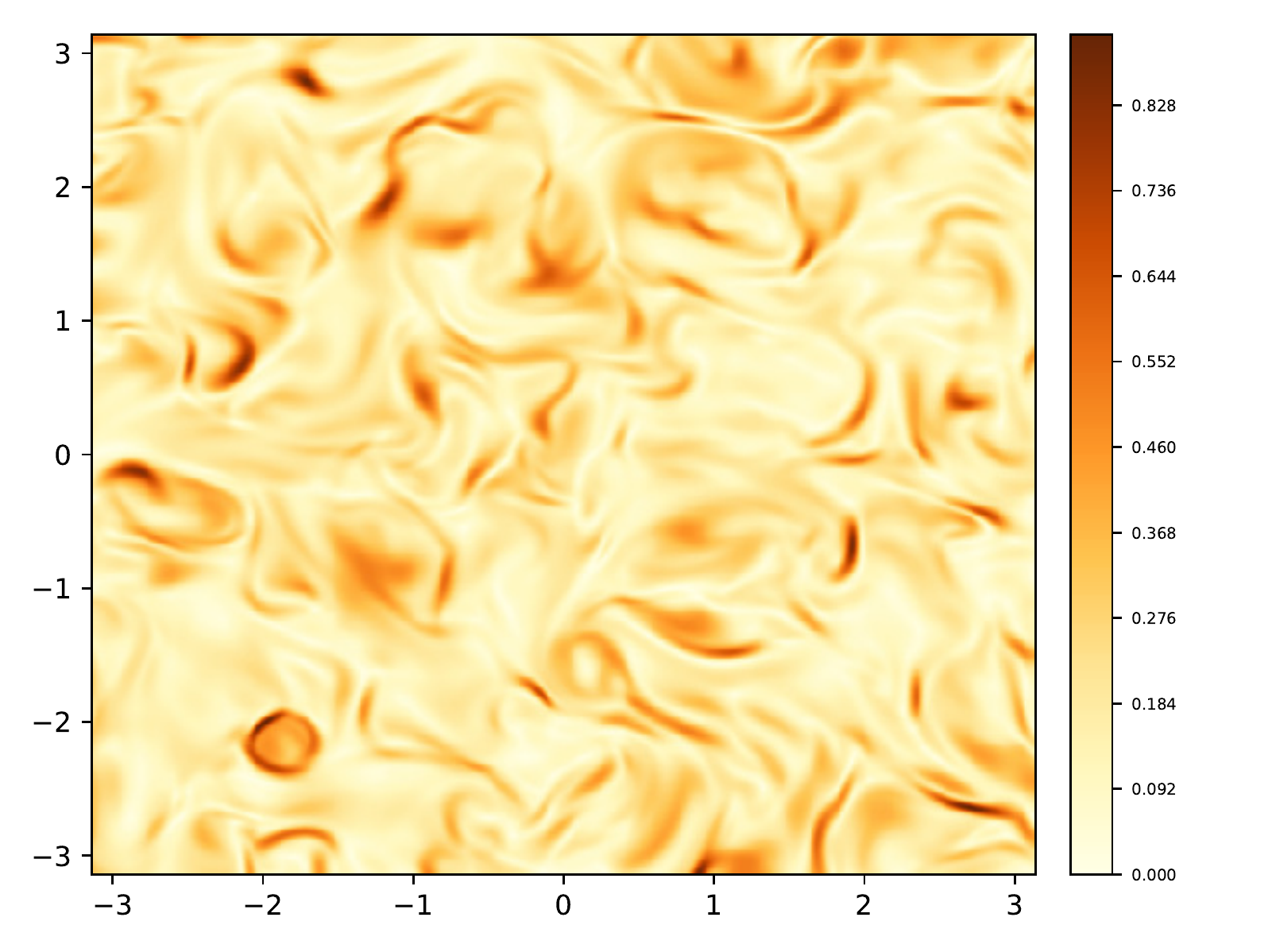}
\caption{
Evolution of the absolute value of the current density ($\vert \JJ \vert$) in 1/4 of the domain of an arbitrary 2D slice from the 3D simulation, A3D, shown in contour plots at times $t=3$, $t=21$ and $t=60$, from left to right.} 
\label{contjevol3d}
\end{figure*}

\begin{figure}
\centering
\includegraphics[width=0.46\textwidth, height=0.24\textheight]{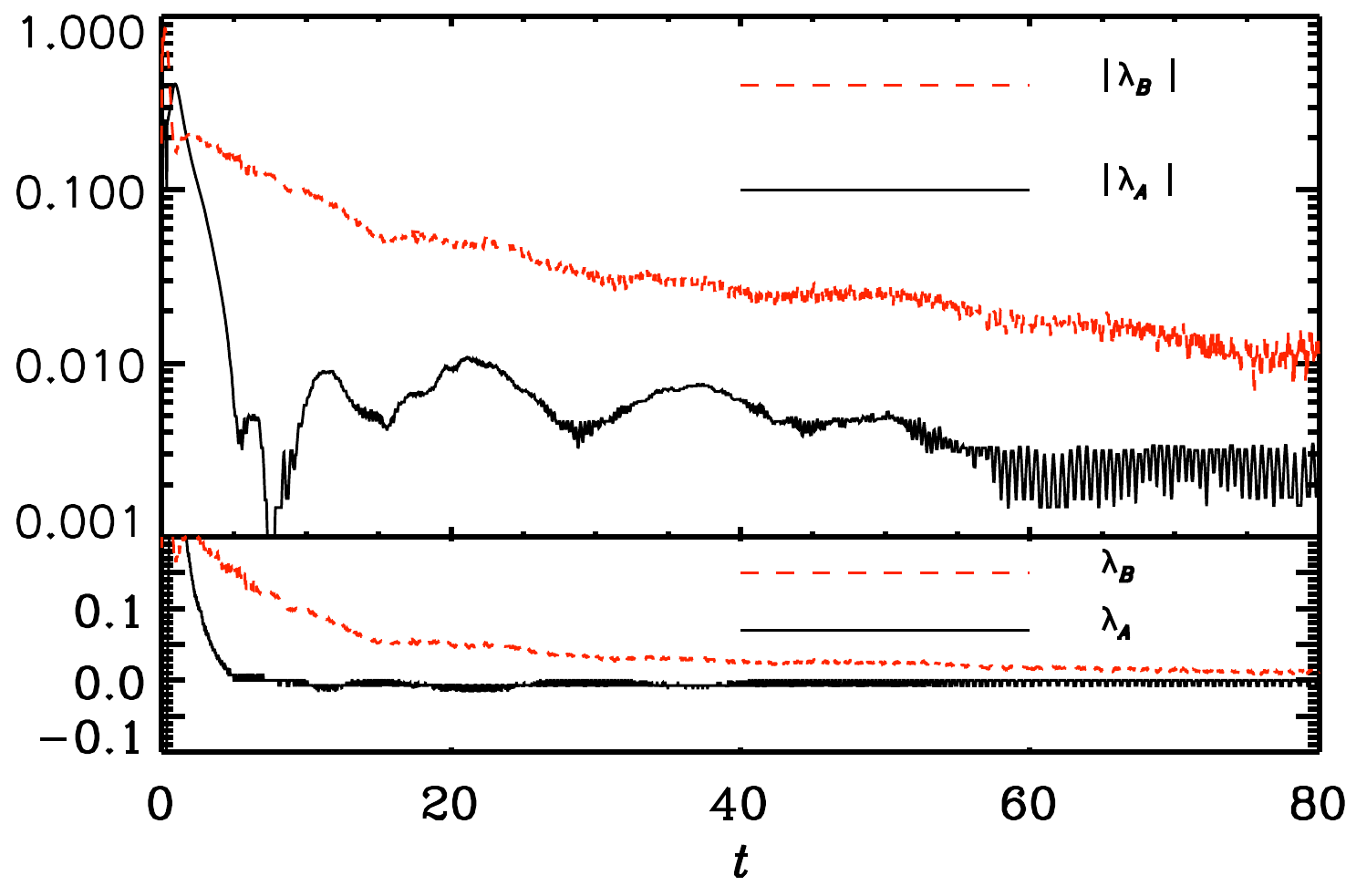}
\includegraphics[width=0.46\textwidth, height=0.24\textheight]{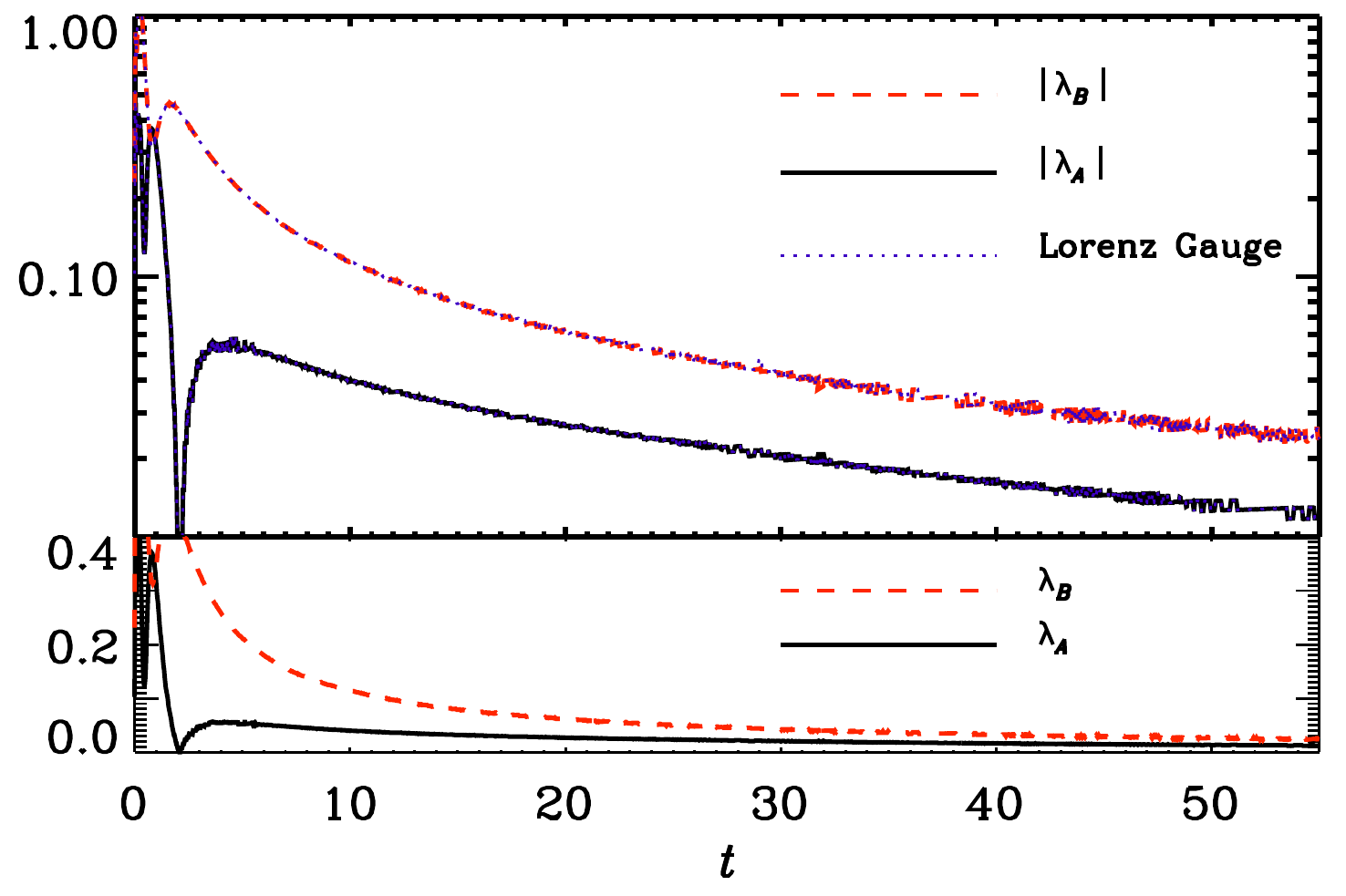}
\caption{The rate of change of vector potential squared, $\lambda_A$ (dashed red), and magnetic energy, $\lambda_B$ (solid black), 
is shown for 2D and 3D simulations, A2D and A3D (where $S=1000$) in the top and bottom figures, respectively. In each figure, the upper panel is a log-linear plot, 
whereas the lower panel is a linear-linear one. In the bottom figure, 
an additional curve from a 3D simulation employing the Lorenz gauge is shown in dotted blue. 
}
\label{a2cons}
\end{figure}

\begin{figure}
\centering
\includegraphics[width=0.35\textwidth, height=0.17\textheight]{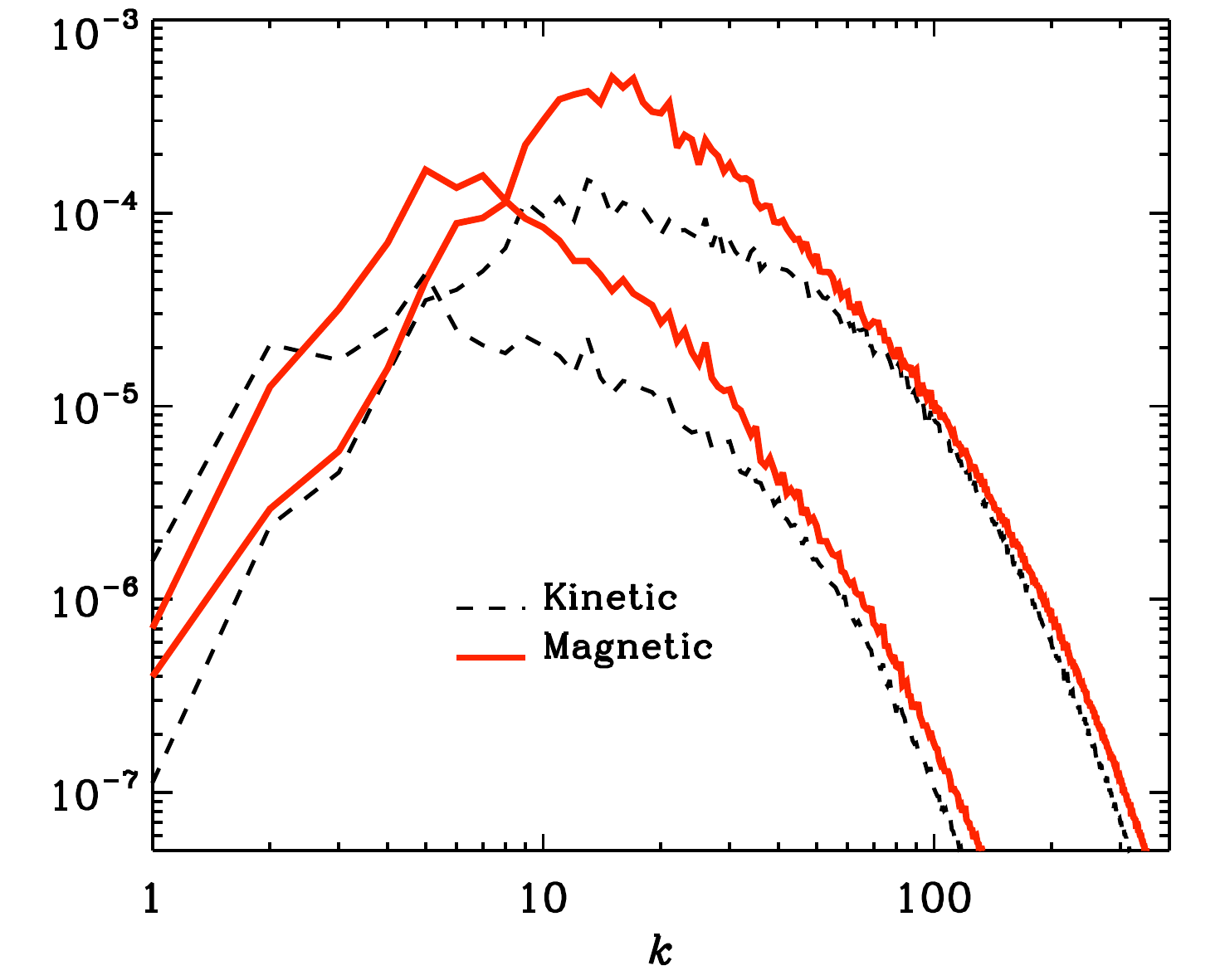}
\includegraphics[width=0.35\textwidth, height=0.17\textheight]{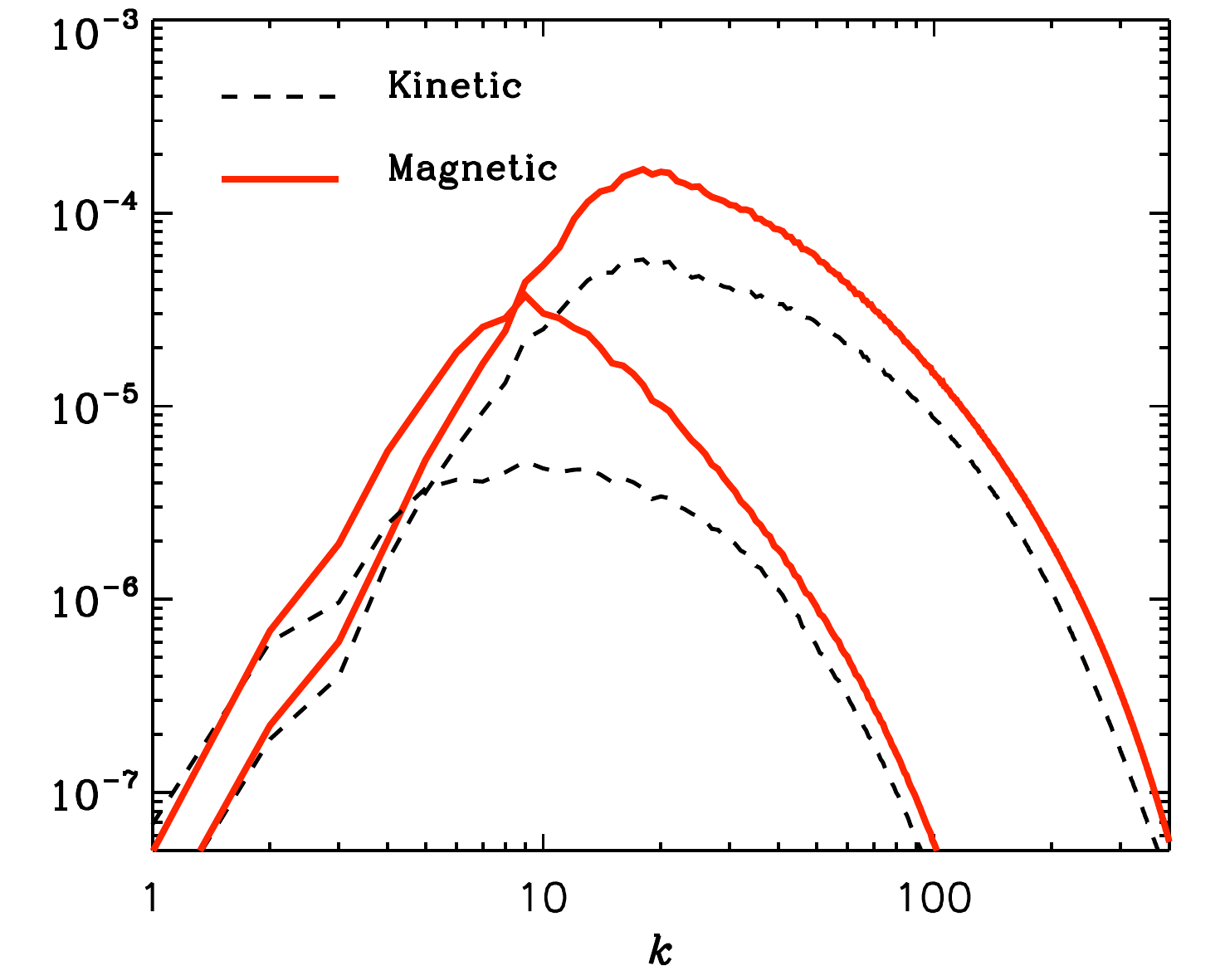}
\caption{
Magnetic and kinetic power spectra plotted at times, $t=2$ and $12$
from the 2D and 3D runs, A2D and A3D, in upper and lower panels respectively.}
\label{spectra2}
\end{figure}

Next, we turn to 3D simulations (runs A3D, B3D, C3D, D3D and E3D in Table~\ref{xxx}). The 3D run resolutions go up to $1024^3$ grid points, and all have an initial condition similar 
to the 2D case of random magnetic fields with power peaked at small scales, as specified in \Eq{initcond}. 

As in the 2D case, we  again observe a power-law-in-time magnetic energy decay  with exponent $-1$, as shown
in the top panel of \Fig{3Devol} (at later times, the decay of magnetic energy steepens, possibly due to diffusion beginning to dominate the system. \citet{axel2015}, who use a similar setup, do not report such a transition, possibly because the higher resolution that they employ ($2304^3$) reduces diffusive effects in their simulation).

From the bottom panel of \Fig{3Devol}, a flat range indicating $k^{-2}$ slope in the magnetic spectrum 
can be observed \citep{axel2015,zrake2014}, however with limited range given the resolution.  
These scalings are intriguingly similar to the ones seen already in the 2D case, thus triggering the following questions: 
\begin{enumerate}
\item To what extent are the 3D simulations similar to the 2D ones? 
\item Can we conclude that, even in 3D, "structure mergers" via reconnection are responsible for this inverse transfer of magnetic energy?
\end{enumerate}
This section and the next are concerned with answering these questions. 

Firstly, we see in \Fig{contevol3d} qualitative similarities with the 2D runs; namely, the evolution of the magnetic field structures 
(from a slice out of the 3D domain) resembles the behaviour of the magnetic islands seen in the contour plots from the 2D system (\Fig{contevol}). 
We also show the evolution of the $x$-component of the field, $B_x$, in \Fig{boxevol3d}. 
It is clearly seen that the field structures grow in scale. However, here in the 3D case, the structures are more elongated and are not as symmetric as in the 2D case.  
Nonetheless, they do not exhibit any specific directionality overall.
In other words, while locally each field structure does seem to prefer 
a certain direction (given the elongation), these preferences are randomly distributed over the domain. Thus, there is no development of a large-scale structure which can bias the system in a certain randomly chosen direction, as is routinely seen, for example, in helical dynamos \citep{BS05}. 

From comparisons with the 2D results, there is a suggestion that perhaps, even in the 3D system, a reconnection-based mechanism might be responsible for the growth of the structures over time
(\citet{munietal2020} have explored the suggestion in this work, also in the context of reduced-MHD, and found it to correctly describe their numerical results).
To further support this suggestion, we show in \Fig{contjevol3d} the absolute value of the current density, $\vert \JJ \vert = \sqrt{J_x^2+J_y^2+J_z^2}$). The wispiness of the current density structures corroborates the existence of current sheets where reconnection can take place.

Already at this point it is possible to argue for why there are similarities between the 2D and the 3D results. Given that the system is magnetically dominated, we think that strong local anisotropy spontaneously arises. This is reflected in the previously mentioned elongation of the magnetic structures in \Figs{contevol3d}{boxevol3d}. This local anisotropy, then, could lead to 2D-like behaviour. While the Sweet-Parker scaling of magnetic reconnection rate does not change from 2D to 3D, this local anisotropy entails the possibility of the existence of local guide-fields, if required to render the 3D reconnections with 2D-like behaviour. This would explain why we see results in 3D which are similar to that in 2D (such as the magnetic energy scaling of $t^{-1}$ and the spectral scaling of $k^{-2}$). 

Next, we look at the conservation properties in both the 2D and 3D systems. First, we show in the top panel of \Fig{a2cons} 
the evolution of the rate of change of the 2D MHD ideal invariants $P$ (black) and $\EEM$ (red-dashed) (given that kinetic energy is subdominant here), 
$\lambda_A=d(\ln{P})/dt$ and $\lambda_B=d(\ln{\EEM})/dt$, respectively, calculated from run A2D. 
As expected, $\lambda_A$ is much smaller than $\lambda_B$, thus demonstrating $P$ to be better conserved than $\EEM$, as we have argued earlier. 
In the bottom panel of \Fig{a2cons}, we show the evolution of $\lambda_A$ and $\lambda_B$ from the 3D simulation A3D, and 
again we find the former to be much smaller than the latter. 
While theoretically $P$ is strictly an ideal invariant only in 2D, these results suggest that it is possible to make a case for its approximate conservation in 3D as well. 

Consider, therefore, the evolution of $P$ in 3D, 
\EQ
\int dV~ \frac{D}{Dt} \left(\frac{\AAA^2}{2} \right)  = \int dV~\uu \cdot \left( \AAA \cdot \nab \AAA \right) -\eta \BB^2. 
\label{acons3d}
\EN
This equation differs from the 2D case only by the term $\uu \cdot \left( \AAA \cdot \nab \AAA
\right)$ on the RHS.  Here, again, we appeal to the fact that flow is subdominant to the field
in order to assume that backreaction of the flow on the field is negligible. 
Such subdominance can be seen in \Fig{spectra2}: in the vicinity of the peak wavenumber, the amplitude of the kinetic power spectra are lower than the  magnetic power spectra by about an
order of magnitude, in both 2D and 3D cases.  Furthermore, the source term $\uu \cdot \left( \AAA \cdot
\nab \AAA \right)$ in question from \Eq{acons3d} can be compared to the analogous source term in the
equation for the magnetic energy \Eq{cons1}, $\uu \cdot \left( \BB \cdot \nab \BB \right)$ (Note that \Eq{cons1} is valid in 3D also). This term arises on expanding
$\uu\cdot (\JJ \times \BB)=\uu \cdot(-\nab (\BB^2/2) + \BB \cdot \nab \BB)$.
Assuming  that $\lvert\BB\rvert \sim k_{\rm corr} \lvert\AAA\rvert$ (in a scenario where most of the power is in a single scale, represented by the wavenumber $k_{\rm corr}\gg 1$), then these sources  differ by a
factor of $k_{\rm corr}^2$, with the term $\uu \cdot \left( \AAA \cdot \nab \AAA \right)$ being smaller
of the two. Thus, again, we conclude that in the limit of $\eta \to 0$, $P$ decays much slower
than $\EEM$.
Consequently, it follows from \Eq{kcorr} that there can be an inverse transfer in 3D as well, as seen in the 3D simulations. 
\begin{figure}
\centering
\includegraphics[width=0.48\textwidth, height=0.42\textheight]{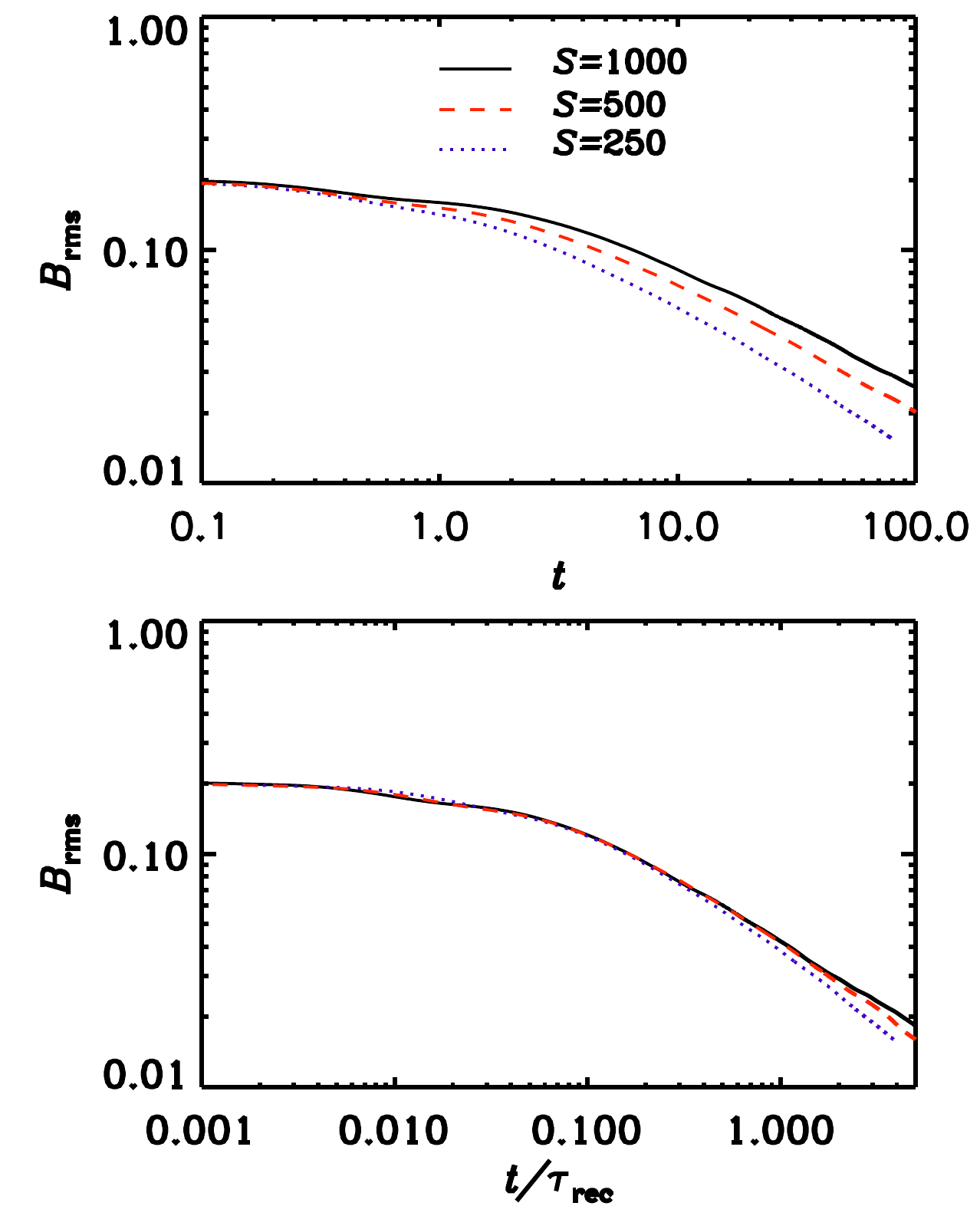}
\caption{Time evolution of $\Brms$ from 2D runs with varying values of $S$. In the lower panel, the time axis has been normalized by the reconnection
timescale $\tau_{rec}$ pertaining to each value of $S$.
}
\label{Sscaling2D}
\end{figure}
\begin{figure}
\centering
\includegraphics[width=0.48\textwidth, height=0.42\textheight]{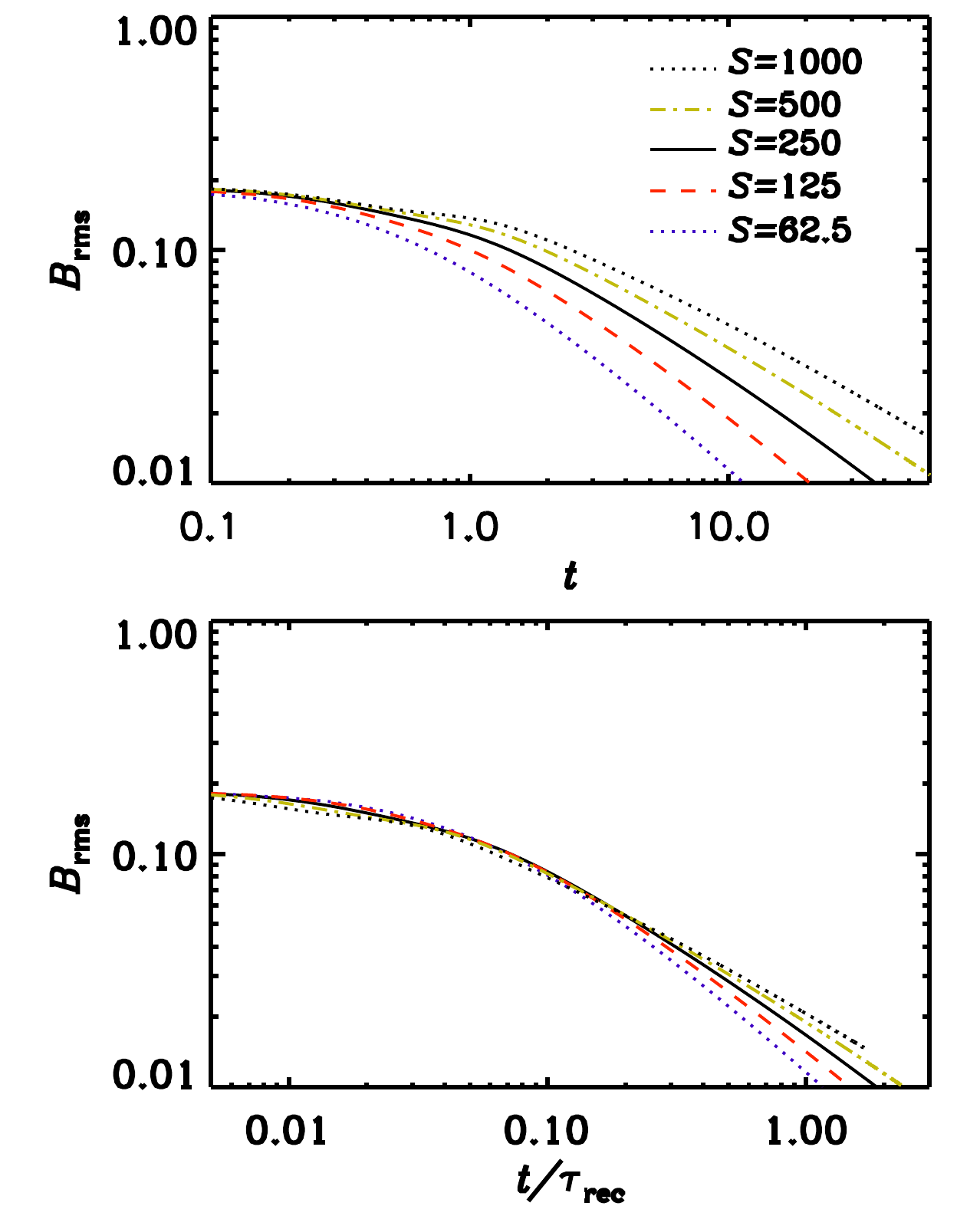}
\caption{Time evolution of $\Brms$ from 3D runs with varying values of $S$. 
In the lower panel, the time axis has been normalized by the reconnection
timescale $\tau_{rec}$.
}
\label{Sscaling3D}
\end{figure}

Since we are dealing with quantities based on vector potential, a fair concern is with regard to the
gauge dependence. As mentioned earlier, our model equations adopt the Weyl gauge ($\Phi=0$). To check for possible gauge-related effects in the results, we performed a simulation using instead the Lorenz gauge, with the same
parameters and initial conditions as those employed in our main runs with the Weyl Gauge. In the Lorenz gauge (or the
pseudo-Lorenz gauge), we have $\partial_t \Phi = - c_{s}^2 \nab \cdot \AAA$
\citep{axelkapyla2007}, where $c_{s}$ is the speed of sound instead of the speed of light.  We
overplot the result in the bottom panel of \Fig{a2cons}  (dotted blue line).  It can be seen that
the results from the Lorenz gauge are indistinguishable from those with the Weyl gauge. This is consistent with the expectation of better conservation of $P$ than of $\EEM$ to hold up in any gauge within a
closed domain, as the sink terms in the equations for $P$ and $\EEM$ remain the same.

While these arguments based on ideal conserved quantities are useful to provide plausibility to the
notion that the understanding of 3D nonhelical inverse transfer lies in its 2D like behaviour, we still do not have
more substantial evidence for reconnection being the driving factor for the inverse transfer. 
To gain a better understanding of the system, we study the timescale governing its dynamical evolution. In doing
so, we continue to probe 
the similarities between the 2D and 3D cases.

The power law governing the evolution of the magnetic field in the 2D system is expected to be $\Brms=B_0
(t/\tau_{rec})^{-1/2}$ as shown by \cite{Munietal2019}, 
where $\tau_{rec}$ is the reconnection time scale, given by
$\tau_{rec}=\beta_{rec}^{-1}(2\pi/k_{\rm corr0})/V_{A0}$, with $\beta_{rec}$ the normalized reconnection
rate, $k_{\rm corr0}$ is the wavenumber associated with the initial correlation scale and $V_{A0}$ is
the initial Alfv\'en velocity. Here we use the Sweet-Parker scaling for the reconnection rate \cite{Sweet1958,Parker1957},
$\beta_{rec}=S^{-1/2}$, which is appropriate for values of $S$ lower than the value critical to
trigger the plasmoid instability \citep{nunoetal2007,samtaney_2009}.  Note that as the simulation proceeds, the correlation
scale, $(2\pi/k_{\rm corr})$ (we take $k_{\rm corr}=k_p$) increases and the Alfv\'en velocity, $V_A$,
decreases, and thus the Lundquist number, $S=V_A(2\pi/k_{\rm corr})/\eta$ is expected to remain constant
\citep{Munietal2019}. For two different runs with different Lundquist numbers $S_1$ and $S_2$, at
any given time $t$, the ratio of the magnetic field strengths is then predicted to scale as
$B_{\rm rms1}/B_{\rm rms2}=(S_1/S_2)^{1/4}$.

In \Figs{Sscaling2D}{Sscaling3D}, we compare $\Brms$ evolution curves from 2D and 3D runs,
respectively, with different values of $S$, which vary by a factor of 2 from one run to another.  In
the bottom panels of \Figs{Sscaling2D}{Sscaling3D}, we normalize the time axis by the 
reconnection timescale $\tau_{rec}$ (note that the normalization $\tau_{rec}$ is computed for the initial $k_{\rm corr}$ and not varied with time; this is because $k_{\rm corr}$ is a discrete quantity and thus its variation does not lead to a secular evolution of the time axis $t/\tau_{rec}$). 
On applying this normalization, there is a notable tendency for curves from different simulations to 
collapse on top of each other.
The collapse of the curves is better in the 2D case than the 3D case;
but, even in the 3D case, for runs with increasing values of $S$, the gap between the successive curves decreases. The curves from runs with
the highest resolution and Lundquist numbers, $S=500$, shown in dash-dotted green, and $S=1000$, shown
in dotted black, very nearly collapse on top of each other. 
These results  
suggest that the reconnection timescale dictates the dynamical evolution of both the 2D and the 3D systems.
A point to be noted is that when the time axes are normalized by the resistive timescale instead, the curves do not collapse together. 

This result of curves collapsing together on normalization of time by $\tau_{rec}$ strongly supports the
possibility of magnetic reconnection being the key mechanism responsible for this 3D nonhelical
inverse transfer.

\subsection{Energy transfer functions}
\label{tfrfunc}

The previous sections have provided both qualitative and quantitative information in support of the notion that magnetic reconnection is the physical mechanism underlying the inverse transfer that we observe in both the 2D and 3D simulations. 
Additional arguments consistent with this conclusion arise from the analysis of the energy transfer functions, as we discuss in this section.

We calculate spectral transfer functions involving transfer 
between different scales in the magnetic energy, given by $T_{\bm{bb}}$, between magnetic and kinetic energies, 
given by $T_{\bm{ub}}$, and between different scales in the kinetic energy, given by $T_{\bm{uu}}$. For the calculation of the transfer functions, we follow the formalism discussed by \citet{Greteetal17}.
The transfer function $T_{xy}(Q,K)$ denotes the transfer of energy from shell $Q$ to shell $K$, with the subscript 
refering to the energy reservoir, $u$ for kinetic energy and $b$ for magnetic energy. In other words, $T_{xy}(Q,K) > 0$ denotes a transfer from the reservoir $x$ 
to $y$, and $T_{xy}(Q,K) < 0$ denotes transfer from $y$ to $x$. These functions are antisymmetric when $x=y$,i.e., $T_{\bm{uu}}$ and $T_{\bm{bb}}$. The transfer functions are given by
\begin{eqnarray}
T_{\bm{bb}}(Q,K) &=& - \int \BB^K \cdot \left( \uu \cdot \nab \right) \BB^{Q} \nonumber \\ 
&+& \frac{1}{2} \BB^K \cdot \BB^{Q} \left( \nab \cdot \uu \right) d\xx, \\
\label{tfrbb}
T_{\bm{ub}}(Q,K) &=& \int \BB^K \cdot \nab \cdot \left( \frac{\BB}{\sqrt{\rho}} \otimes \ww^Q \right) \nonumber \\ 
&-& \BB^K \cdot \BB \nab \cdot \left(\frac{\ww^Q}{2\sqrt{\rho}}\right)d\xx, \\
\label{tfrub}
T_{\bm{uu}}(Q,K) &=& - \int \ww^K \cdot \left( \uu \cdot \nab \right) \ww^{Q} \nonumber \\ 
&+& \frac{1}{2} \ww^K \cdot \ww^{Q} \left( \nab \cdot \uu \right) d\xx, 
\label{tfruu}
\end{eqnarray}
where $\otimes$ denotes tensor product, $\ww=\sqrt{\rho}\uu$, and the shell-filtered quantities in real space are given by 
$\phi^K(\xx)=\int_K \hat{\phi}(\kk) e^{i\kk\cdot\xx} d\kk$. 

We intend to look for signatures of magnetic reconnection in the transfer function plots calculated from our simulations. Energetically, MHD reconnection  
involves energy transfer from the magnetic to the velocity fields, manifested by the Alfv\'enic outflows along the length of the current sheet that it generates. There is also, in addition, Ohmic dissipation in the current sheet.  

\begin{figure}
\centering
\includegraphics[width=0.48\textwidth, height=0.25\textheight]{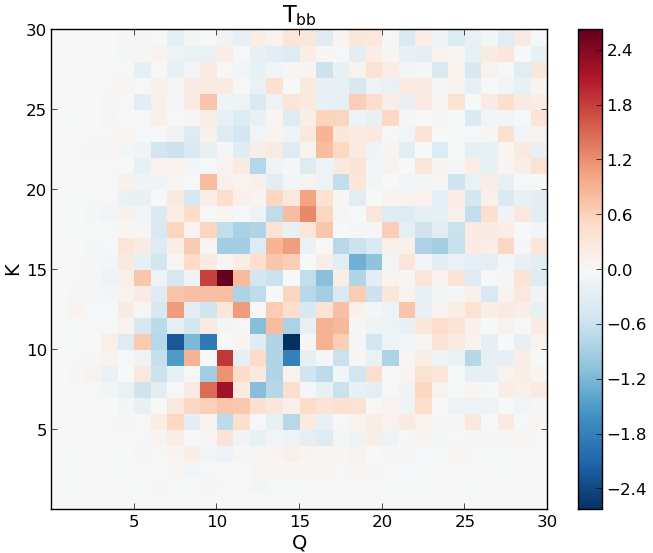}
\includegraphics[width=0.48\textwidth, height=0.25\textheight]{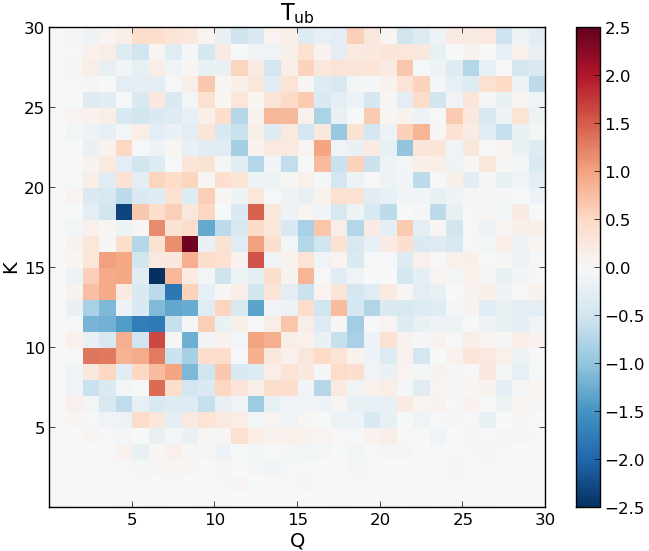}
\includegraphics[width=0.48\textwidth, height=0.25\textheight]{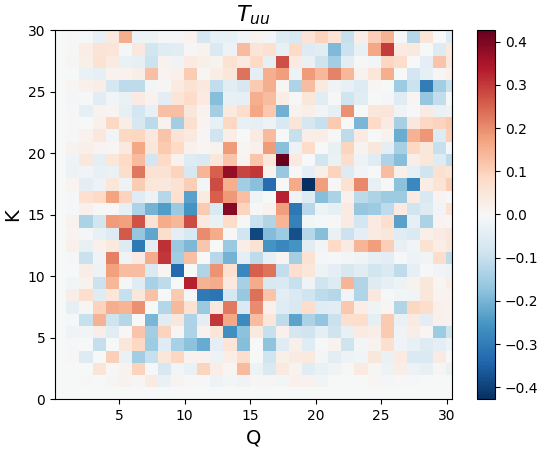}
\caption{
Top, middle and bottom panels show the transfer functions $T_{\bm{bb}}$, $T_{\bm{ub}}$ and $T_{\bm{uu}}$, 
respectively, from the 2D simulation A2D. At this point of time, $t=10$ in the simulation, $k_p\sim 9$.}
\label{2dtfr}
\end{figure}

\begin{figure}
\centering
\includegraphics[width=0.48\textwidth, height=0.25\textheight]{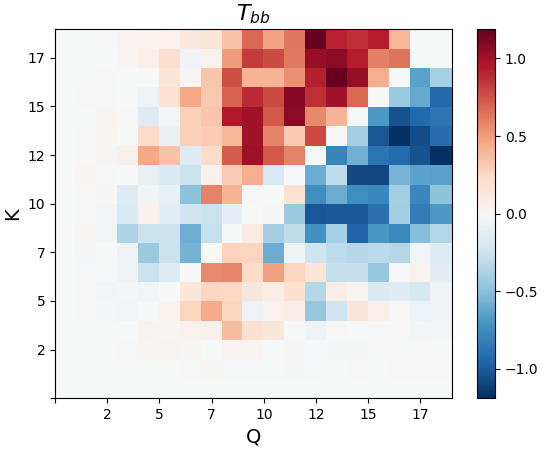}
\includegraphics[width=0.48\textwidth, height=0.25\textheight]{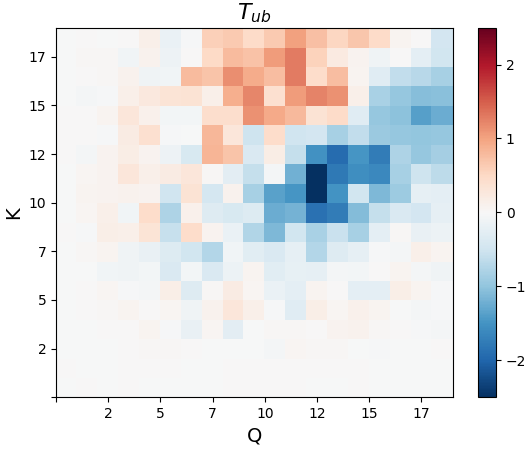}
\includegraphics[width=0.48\textwidth, height=0.25\textheight]{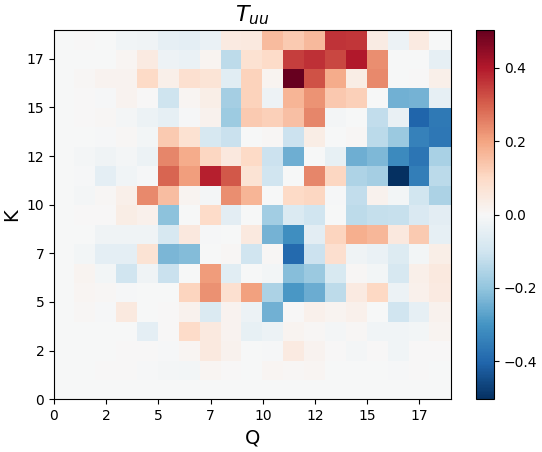}
\caption{
Top, middle and bottom panels show the transfer functions $T_{\bm{bb}}$, $T_{\bm{ub}}$ and $T_{\bm{uu}}$, 
respectively, from the 3D simulation C3D. At this point of time, $t=20$ in the simulation, $k_p\sim 9$.} 
\label{3dtfr}
\end{figure} 
 
In previous sections, we have mentioned that the merging of magnetic islands 
facilitated by reconnection results in inverse transfer in a 2D system; these mergers take place in hierarchical fashion, where each generation of mergers 
produces islands of larger sizes, which then merge to produce larger islands, and so on~\citep{Munietal2019}. We conjecture that the 3D system evolves in a similar way, with reconnection merging current filaments, and resulting in an inverse cascade of magnetic energy.
If this conjecture is true, then we expect to observe, at any given point in time, significant transfer of magnetic to kinetic energy at a scale corresponding to the dominant island size at that time (the current sheet length scales as the size of the islands).

We suppose that the 3D system evolves in a similar way, 
with reconnection merging current filaments, and resulting in an inverse cascade of magnetic energy. 
Given this theoretical understanding, we have the following expectations 
for the transfer function plots:
\begin{enumerate}
\item In the $T_{\bm{bb}}$ plot, the scales at which the merging of islands (or current filaments) predominantly takes place (corresponding to $k_{\rm corr}$) 
should exhibit inverse transfer, while rest of the (smaller) scales should decay or direct transfer to further smaller scales.
\item In the $T_{\bm{ub}}$ plot, the transfer from magnetic to kinetic energy should stand out at scales comparable to those at which the inverse transfer (i.e., reconnection) is dominant.
\item In the $T_{\bm{uu}}$ plot, there should be a similarity with the $T_{\rm bb}$ plot as the flows accompanying the fields will behave similarly. 
\end{enumerate}

Note that the expectations for the behaviour of transfer functions for a system where 
magnetic reconnection drives the inverse transfer are quite specific, as opposed to a case where  
a generic turbulence-related process 
drives the inverse transfer.
For example, in a generic turbulence-related process, we do not expect the transfer from the magnetic to the velocity fields to be concentrated around certain scales, but to be spread out over a wide range of scales. 

In the upper panel of \Fig{2dtfr}, the $T_{\bm{bb}}$ plot from a 2D simulation shows both inverse and direct transfer of energy for certain ranges of scales. Notice that the reflection of the patterns around the diagonal is due to antisymmetry. 
Next, observe that on the lower side of the diagonal, there is a change in the dominant color of red in lower wavenumbers to the dominant color of blue in the higher wavenumbers.
This means that there is inverse transfer of energy from $Q=10$ to $K=6$--$9$ indicated by the red color, and for $Q>10$ forward transfer is dominant, as indicated by the blue color.

In the middle panel of \Fig{2dtfr}, the $T_{\bm{ub}}$ plot shows that energy transfer from the magnetic field 
to the velocity field is from $K=11-14$ to $Q\sim 6$, as indicated by the blue patch. 
Since the blue color refers to negative values, it implies the direction of the transfer 
to be from $K$ to $Q$ and thus from the magnetic to the kinetic energy reservoirs. This confirms that the transfer is localized to a certain set of scales as expected for a phenomenon (reconnection) dependent process as opposed to a generic turbulence driven process. 

The bottom panel of \Fig{2dtfr} shows the $T_{\bm{uu}}$ plot. Below the diagonal line, the darkest red spot at $Q=10$ and the surrounding small red patch is indicative of minor inverse transfer of energy. This is mainly due to the reasoning that as the magnetic structures merge, the underlying flow structures also acquire a larger size. In that sense, the features in $T_{\bm{uu}}$ plot mimic the $T_{\bm{bb}}$ plot. Also, given that the flow is energetically subdominant to the field, the $T_{\bm{uu}}$ transfers are expected to be small.
\begin{figure}
\centering
\includegraphics[width=0.35\textwidth, height=0.17\textheight]{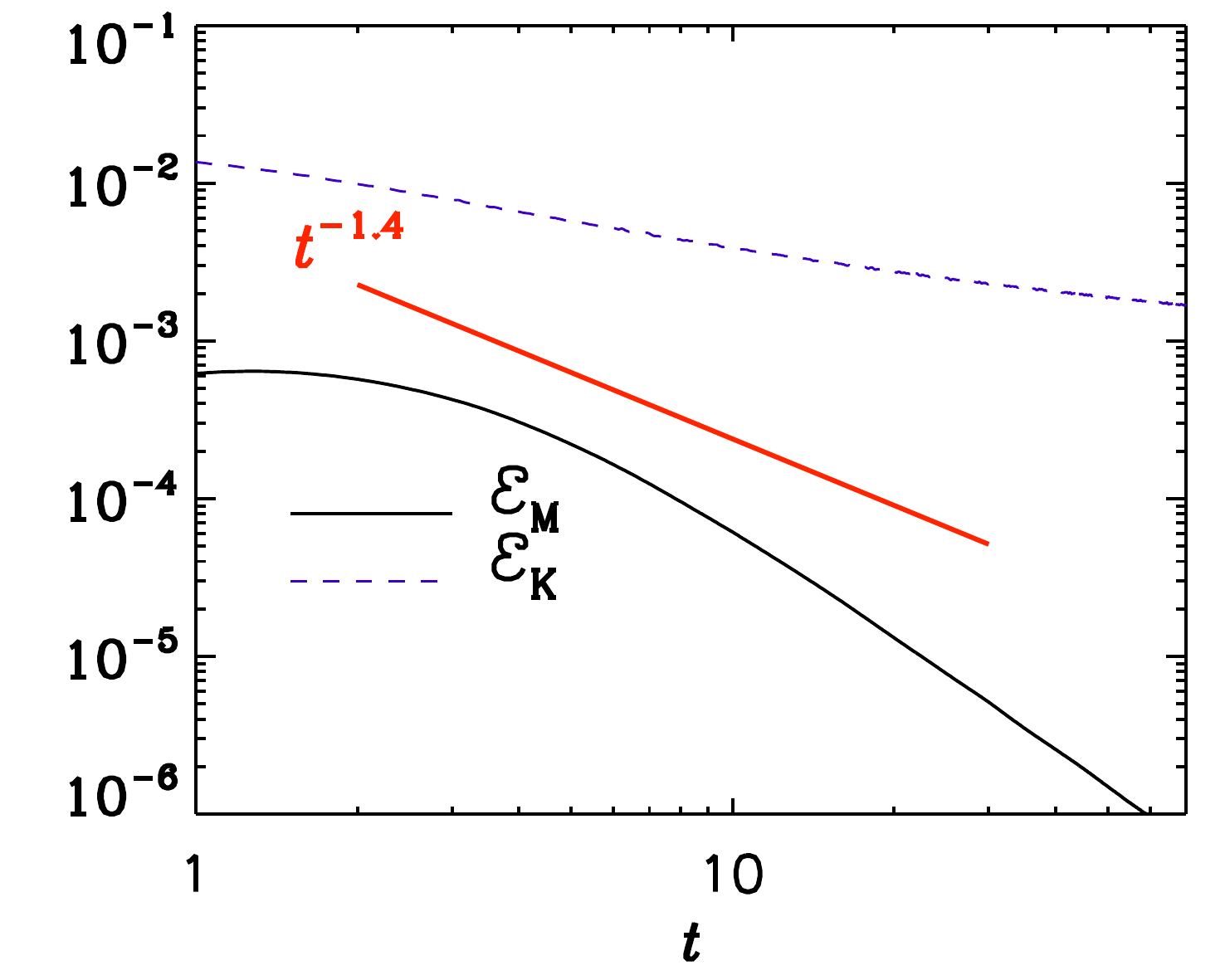}
\includegraphics[width=0.35\textwidth, height=0.17\textheight]{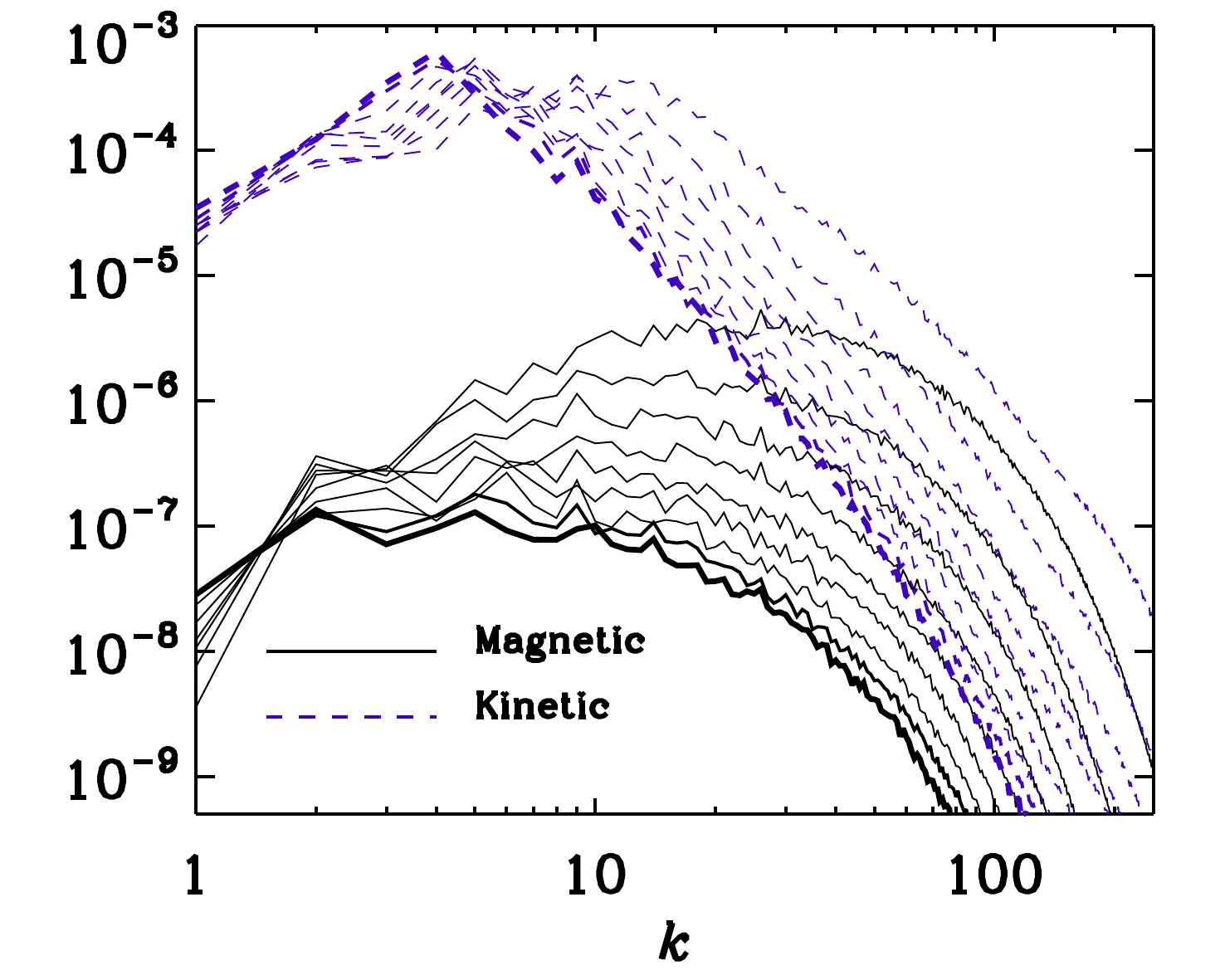}
\caption{
Top panel: Evolution of magnetic energy (solid black) and kinetic energy (dashed blue) in a 2D
simulation, F2D, with non-zero initial velocity.
Bottom panel: Magnetic and kinetic power spectra from the same simulation, plotted at regular intervals of $\Delta t=5$, with a
thick final curve at $t=45$.}
\label{2dinitvel}
\end{figure}

Similarly, we show transfer function plots for the 3D case in \Fig{3dtfr}.
In the plot of $T_{\bm{bb}}$, we find that the pattern changes trend around $Q\sim 10$. 
The scales larger than the wavenumber $Q\simeq 10$ 
exhibit inverse transfer (these are the scales where reconnection would be taking place), while $Q \geq 10$ show forward transfer, as expected. 
In the plot for $T_{\bm{ub}}$, the transfer from magnetic to kinetic reservoirs is localized around $Q\sim 12$ and $K\sim 10$, as expected from a reconnection-dependent process dominantly happening at these scales.
Again, as in 2D case, the $T_{\bm{uu}}$ plot here in 3D shows similarity to the $T_{\bm{bb}}$ plot, with a minor inverse transfer of energy from around $Q=7$.
Note that the energy transfers are mostly local and thus the patterns seen in all the plots are mostly concentrated around the diagonal in both 2D and 3D cases.

The 2D and the 3D transfer function plots tell a similar story --- with greater clarity in the 3D case, we think, because turbulence in that limit is unconstrained. 
The behaviour of the transfer functions is what is expected for a magnetic-reconnection-driven inverse cascade. 

\subsection{The case when the initial velocity field is non-zero}
\label{nonzerov}

In all our simulations up until this point, the velocity field was initialized to be zero.  The flows that arose in these simulations were generated by the magnetic field, and were shown to be subdominant to it.
Magnetic reconnection is typically accompanied by the conversion of magnetic to kinetic energy. These generated flows, thus, are largely Alfv\'enic in nature. And such
flows, where $\uu$ and $\BB$ are mostly parallel, lead to negligible induction.  

However if the velocity field is
non-zero (and the system is not magnetically dominated) to begin with, 
it can lead to a non-trivial stretching term ($\BB\cdot\nab \uu$), resulting
in conversion of kinetic to  magnetic energy.  Then the simple arguments for showing
$\lambda_A \ll \lambda_B$ will not hold true anymore.  This invites the question that if we consider a
non-zero initial velocity field, will we observe energy decay of a different nature, one
without an inverse transfer?  To clarify this question, we have also performed simulations where the initial velocity
field is not only finite, but dominant,  which we discuss in this section.

\begin{figure}
\centering
\includegraphics[width=0.35\textwidth, height=0.17\textheight]{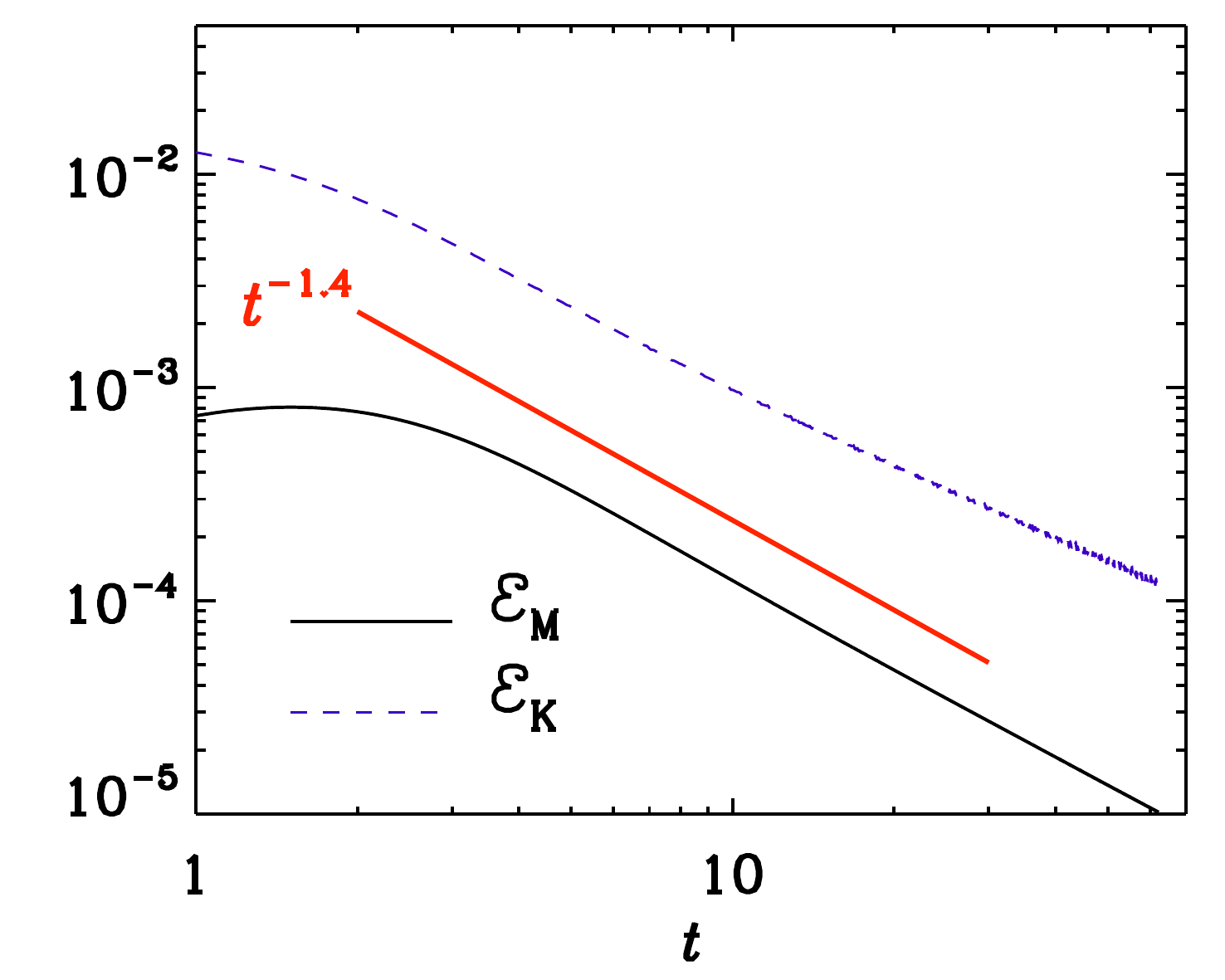}
\includegraphics[width=0.35\textwidth, height=0.17\textheight]{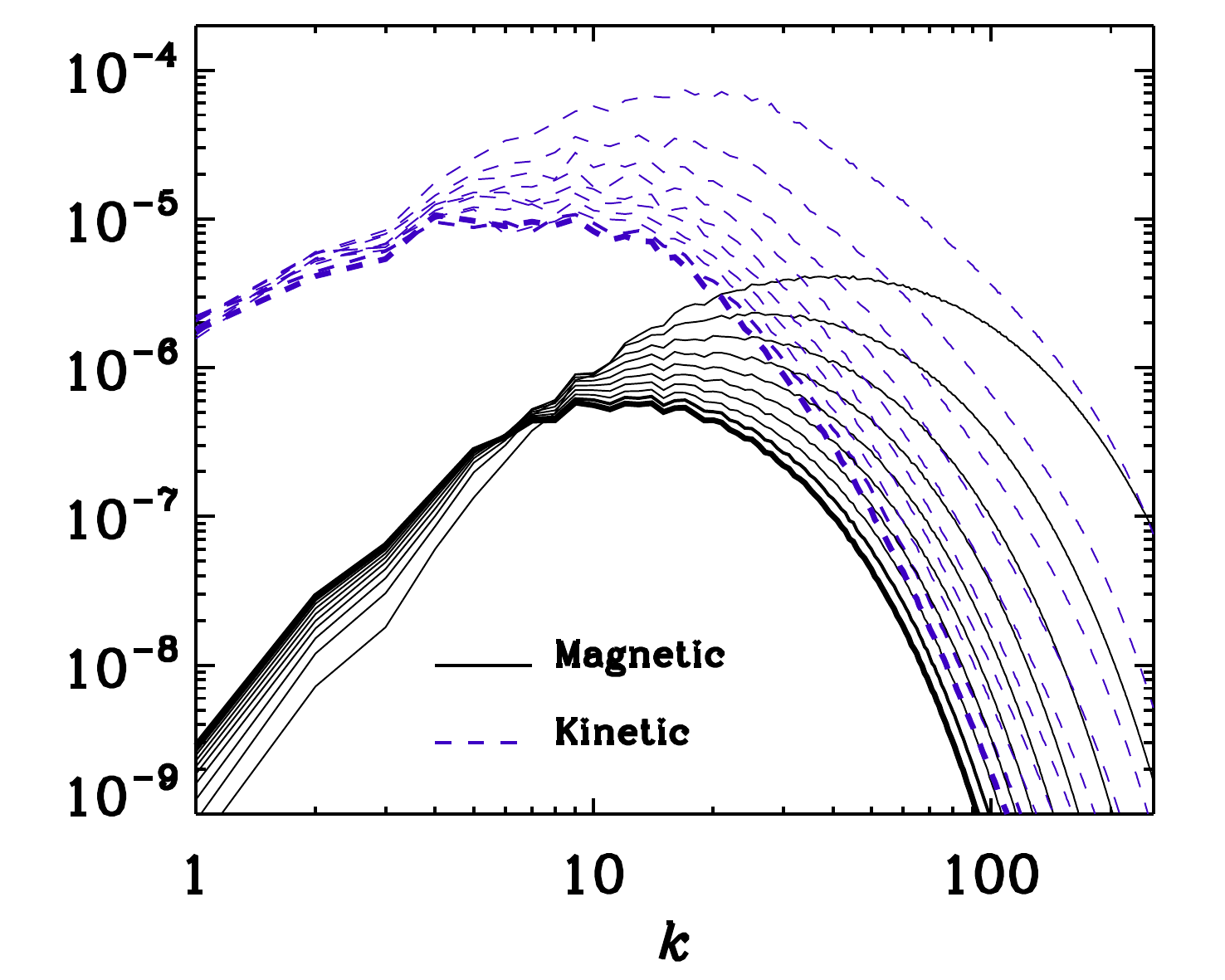}
\caption{
Top panel: Evolution of magnetic energy (solid black) and kinetic energy (dashed blue) in a 3D simulation, F3D, 
with non-zero initial velocity.
Bottom panel: Magnetic and kinetic power spectra $M(k,t)$ plotted at regular intervals of $\Delta t=5$, with a
thick final curve at $t=50$.}
\label{3dinitvel}
\end{figure}

We first examine the 2D case (run F2D). We initialize the flow field in a manner similar to the magnetic field, as
specified in section \ref{sec:ic_and_param}.  While the slope of the magnetic power spectrum is set to $k^4$, the
kinetic spectrum is set to $k^2$ (chosen because this is the slope that develops in the runs when the initial velocity field is zero).  
Also, $\urms$ is initialized to be larger than
$\Brms$ by a factor of 10. In  \Fig{2dinitvel}, we show the evolution curves of the magnetic and kinetic energies,
and also their spectra.  It is seen that there is no inverse transfer in energy (there is a minor growth at the $k=1$ which we will address below), and 
also the
temporal scaling of the magnetic energy evolution curve is much steeper than the $\sim t^{-1}$ evolution found in the case of zero initial velocity (\Fig{2Devol}).

\begin{figure*}
\includegraphics[width=0.33\textwidth, height=0.21\textheight]{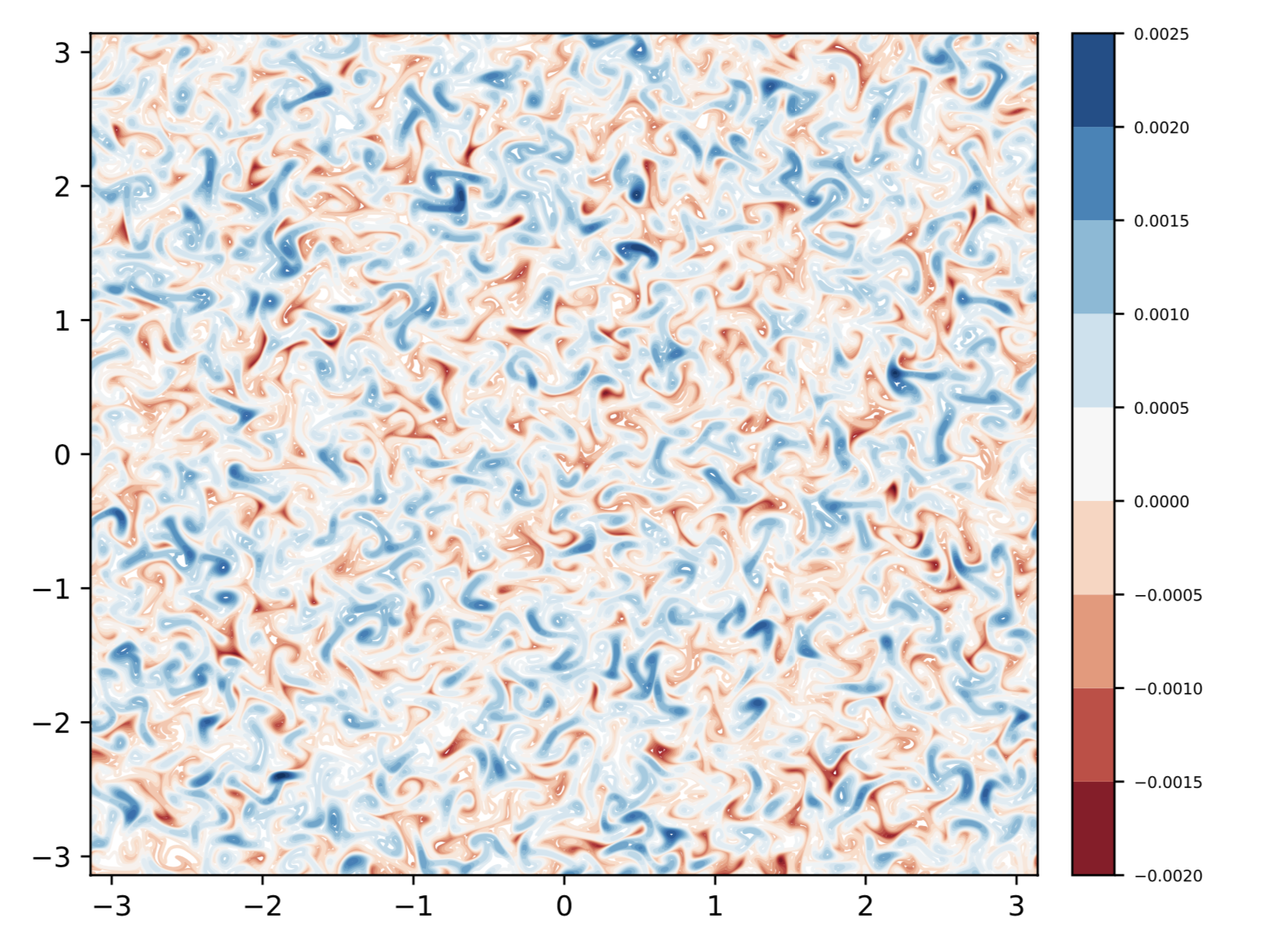}~~
\includegraphics[width=0.33\textwidth, height=0.21\textheight]{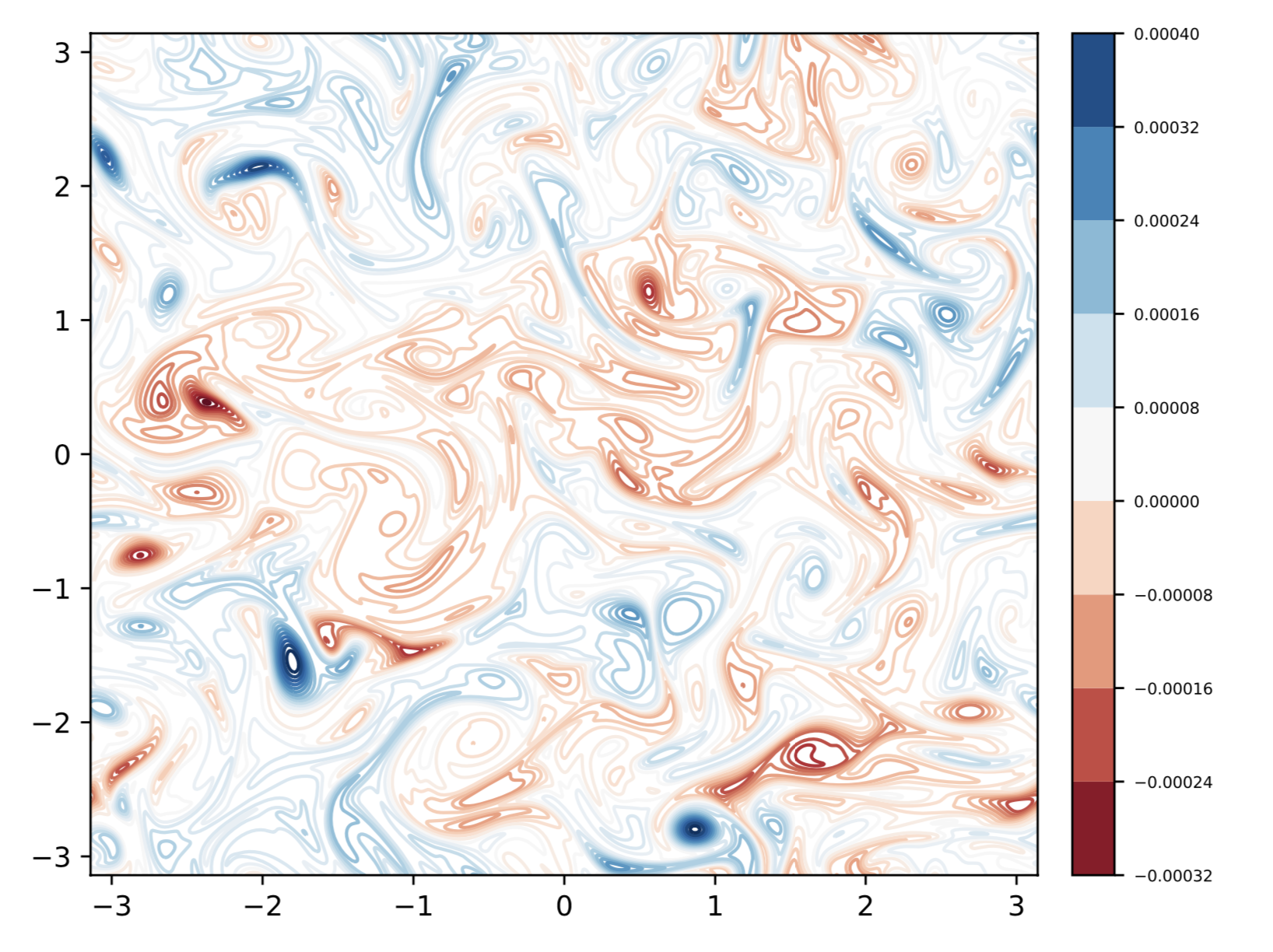}~~
\includegraphics[width=0.33\textwidth, height=0.21\textheight]{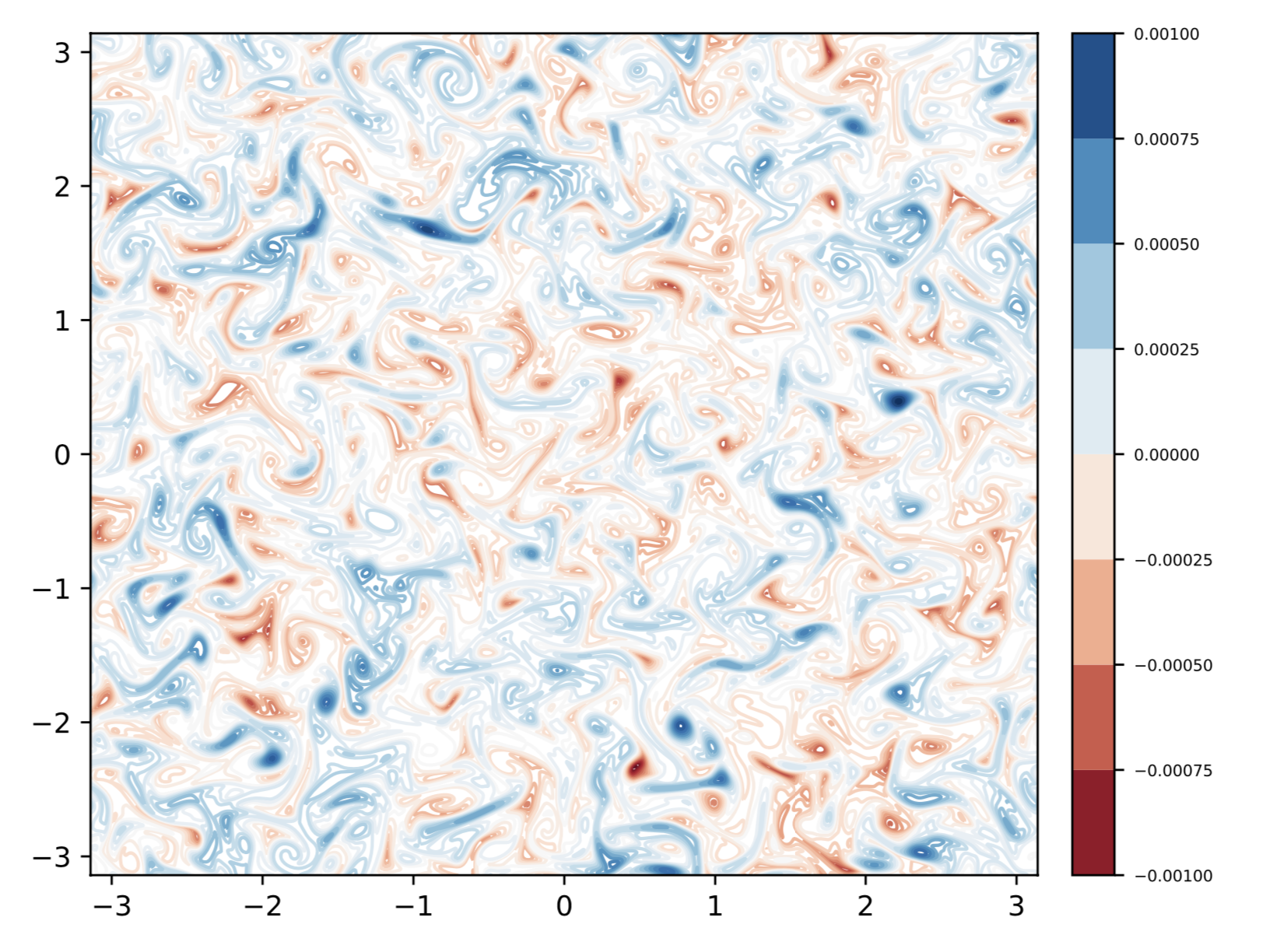}
\caption{Evolution of the vector potential ($A_z$) from the 2D simulation, F2D, with non-zero initial velocity 
shown in contour plots at times $t=2$, $t=10$ and $t=40$ from left to right.}
\label{contevol3}
\end{figure*}
\begin{figure*}
\includegraphics[width=0.33\textwidth, height=0.21\textheight]{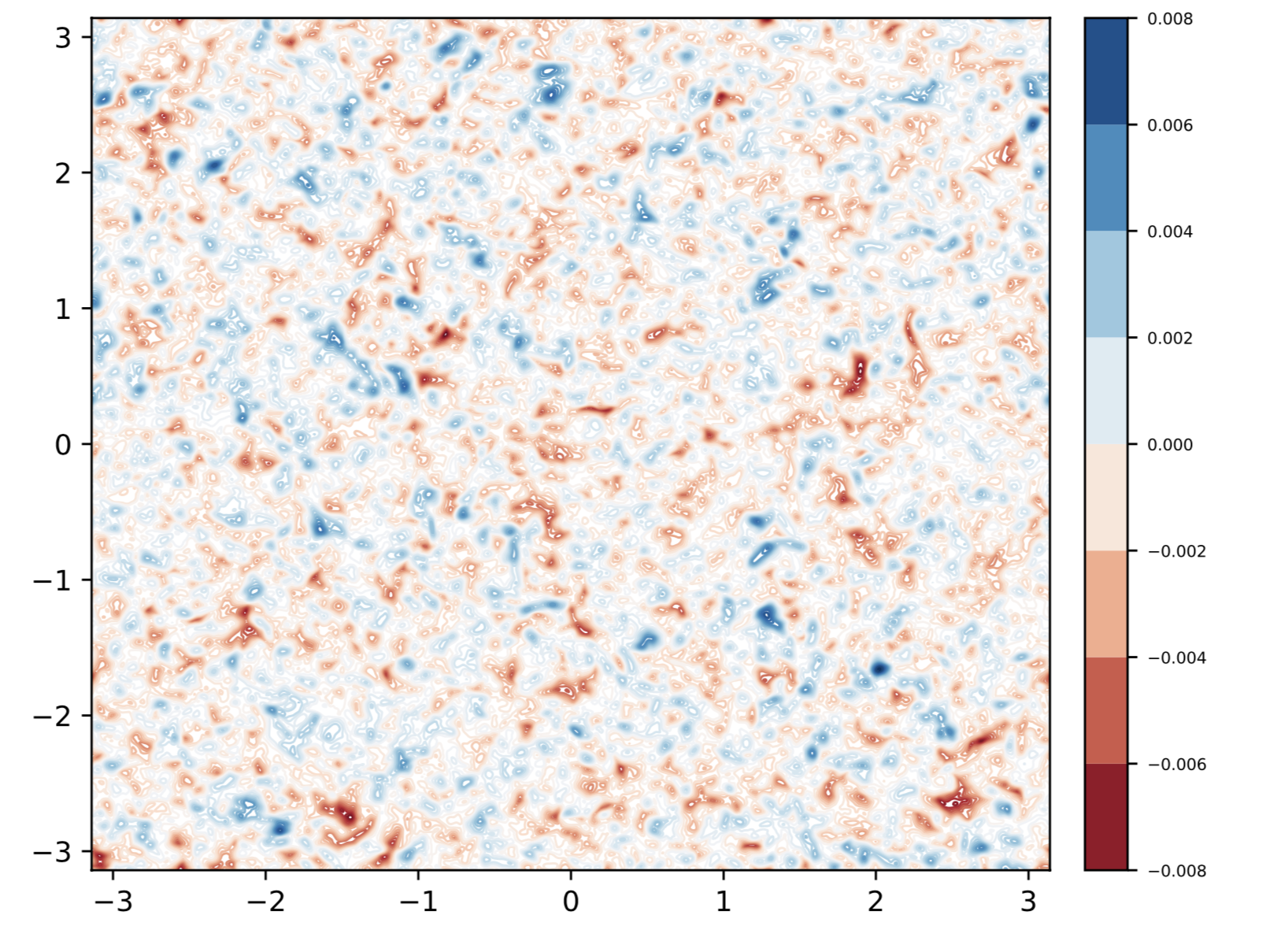}~~
\includegraphics[width=0.33\textwidth, height=0.21\textheight]{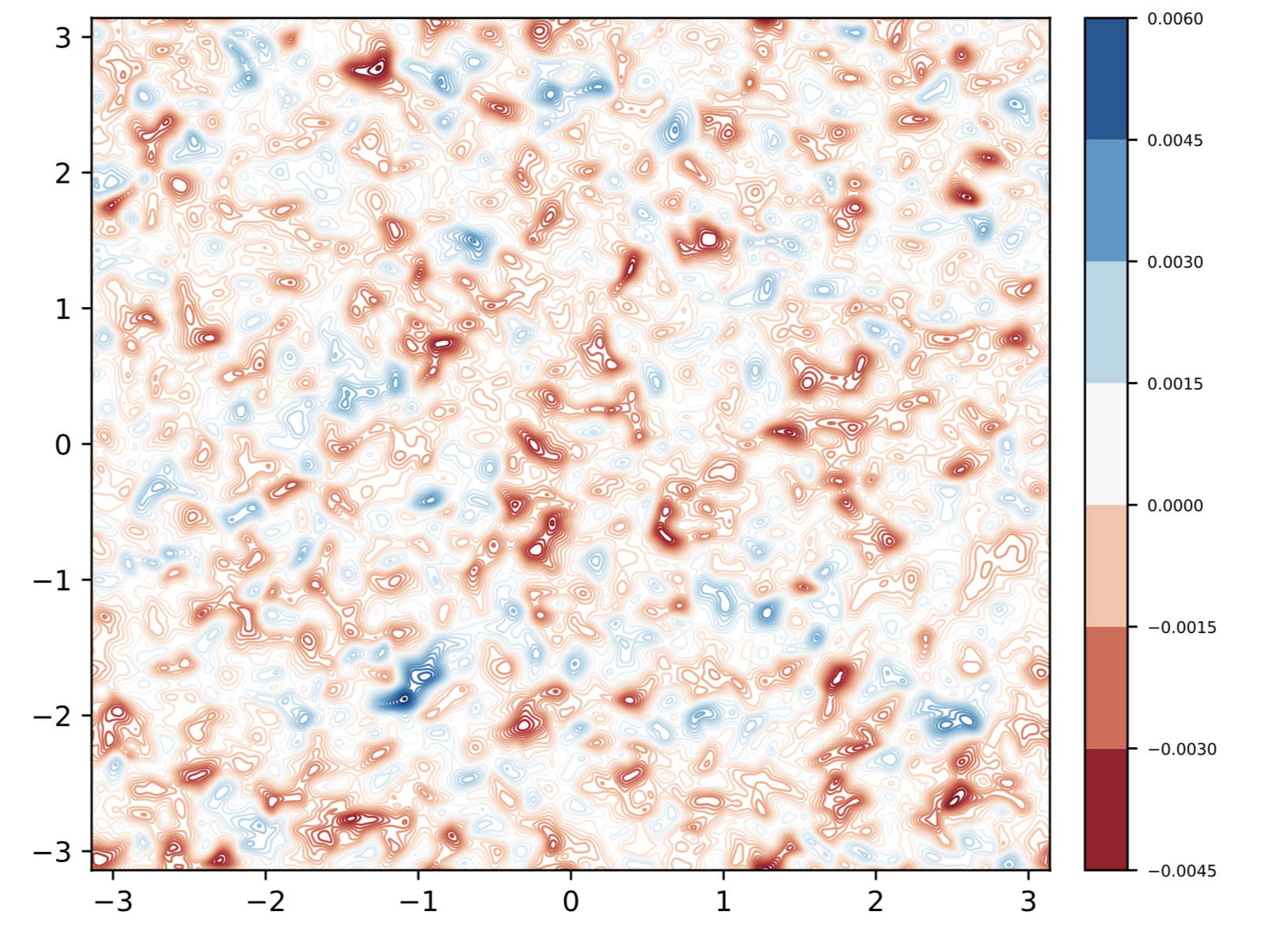}~~
\includegraphics[width=0.33\textwidth, height=0.21\textheight]{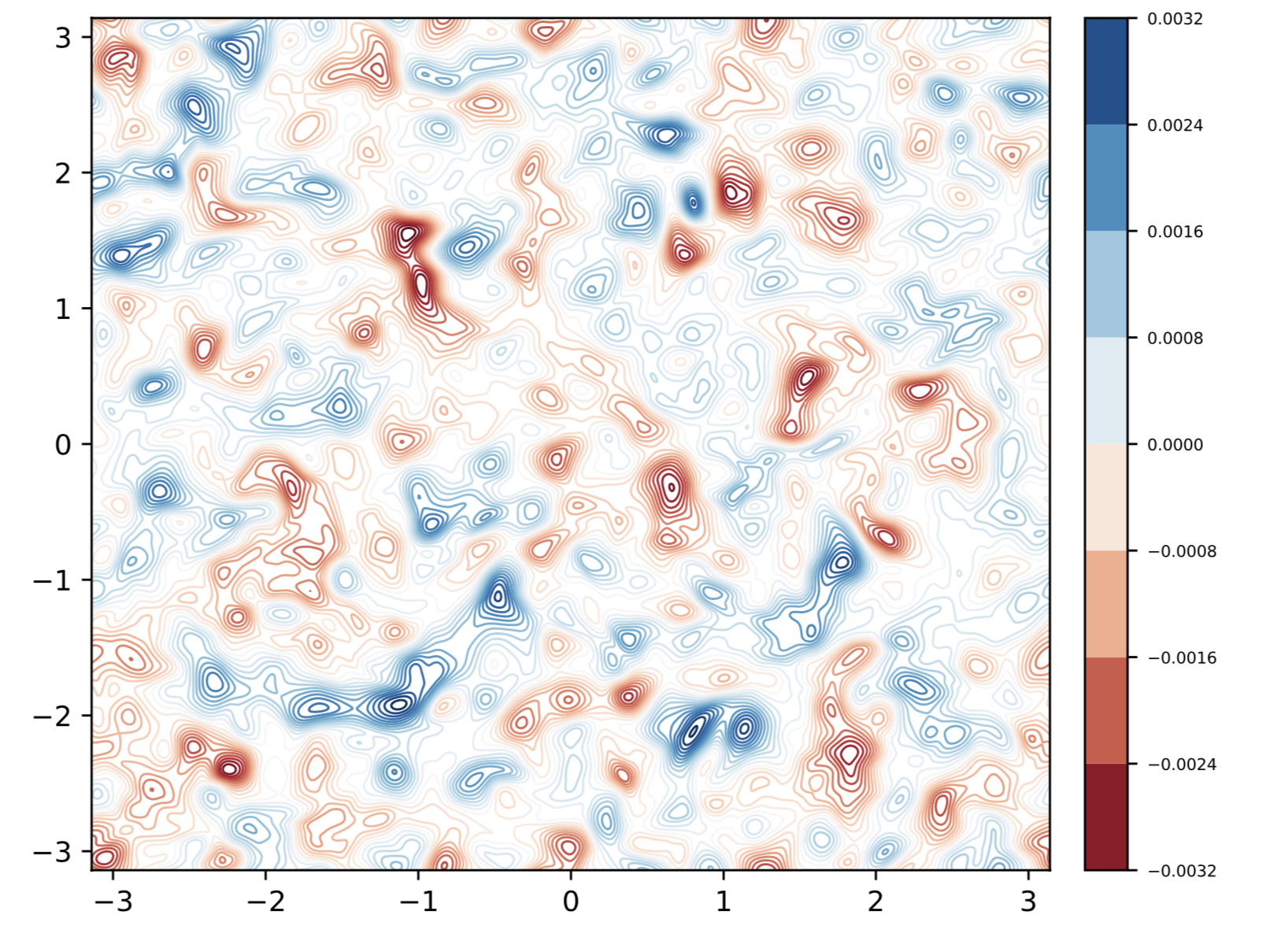}
\caption{Evolution of a component of the vector potential ($A_z$) in an arbitrary 2D slice (in $x-y$ plane) from the 3D domain of the 3D simulation, F3D,  
with non-zero initial velocity shown in contour plots at times $t=5$, $t=20$ and $t=60$ from left to right.}
\label{contevol4}
\end{figure*}

Next, we show in
\Fig{3dinitvel} the evolution curves of the magnetic and kinetic energies,  and their spectra, for the 3D case (run F3D). Here,
surprisingly, we do find an inverse transfer. However the magnetic energy (and the kinetic
energy) does not evolve as $\sim t^{-1}$ but as $\sim t^{-1.4}$.  This numerical scaling of $\sim t^{-1.4}$ is
close to the decay law of $\sim t^{-10/7}$, as governed by the Loitsyanky invariant \citep{davidson} (obtained in the case of hydrodynamic turbulence but not unreasonable to consider here, given the dominance of the kinetic energy).

It is not at once obvious why there is a continued inverse transfer behaviour also when the initial
kinetic energy is non-zero in the 3D case. To understand this, we have to consider that there 
exists a crucial difference between 2D and 3D cases with respect to dynamo action. It is well-known from anti-dynamo theorems  \citep{moffatt,Z57} that there can be no sustained dynamo
action in 2D. A random  velocity field can give rise to anomalous diffusion.
In the absence of any sustained dynamo action, such an anomalous diffusion can lead to rapid decay of
the field in 2D. In \Fig{contevol3}, it can be seen that the system indeed looks turbulent.
The stretching of the fields by turbulence can grow the fields in a certain direction while thinning them out in the perpendicular direction. Thus, even though the structures seem to grow in size over time, they are extremely thin and drawn out. 

In 3D, besides an anomalous diffusion, these same underlying random motions 
can also lead to a dynamo, which can mitigate the effect of the anomalous diffusion. 
The presence of dynamo in our 3D simulations with initial flow can be seen from the top panel of \Fig{3dinitvel}, where the $\Brms$ actually increases 
slightly before it decays. The dynamo effect could explain the difference in the nature of decay 
of magnetic fields in 2D and 3D, when fields are subdominant to random flows.

In \Fig{contevol4}, we find that on a 2D plane from within the 3D domain, 
the magnetic fields structures are not as drawn out as in the 2D case.
They, in fact, retain a more definitive form 
similar to the earlier cases, as in \Figs{contevol}{contevol3d}. 
It is not clear if magnetic reconnection has a role to play in the inverse transfer seen in the 3D case. To investigate this further we now study the transfer function plots obtained for the 3D case.

\begin{figure}
\centering
\includegraphics[width=0.48\textwidth, height=0.25\textheight]{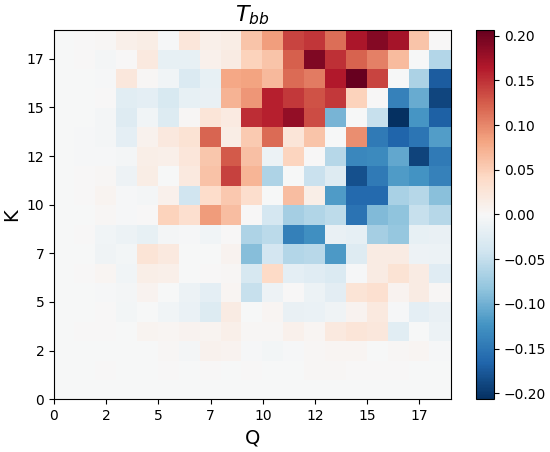}
\includegraphics[width=0.48\textwidth, height=0.25\textheight]{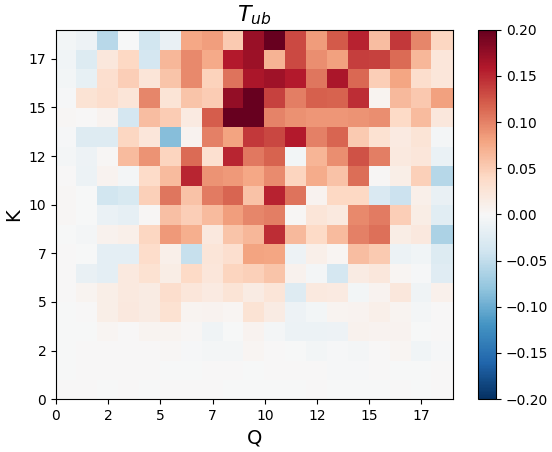}
\includegraphics[width=0.48\textwidth, height=0.25\textheight]{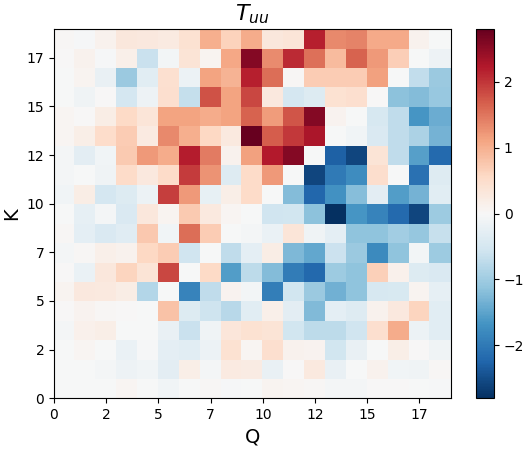}
\caption{
Top, middle and bottom panels show the transfer functions $T_{\bm{bb}}$, $T_{\bm{ub}}$ and $T_{\bm{uu}}$, 
respectively, from the 3D simulation F3D, with non-zero initial velocity. At this point of time, $t=30$ in the simulation, $k_p\sim 9$.} 
\label{nonzerov-tfrfunc}
\end{figure}

\Fig{nonzerov-tfrfunc} presents the energy transfer function plots for the run H3D. Even though the spectra in \Fig{3dinitvel} show the signature of inverse transfer, a corresponding distinctive signature in $T_{\bm{bb}}$ is lacking. The red spots below the diagonal (or equivalently, the blue spots above the diagonal), which indicate inverse transfer, are very few. Here, direct or forward transfer dominates the plot. Also the $T_{\bm{ub}}$ plot is dominated by red color, indicating that the transfers are from kinetic to magnetic energy, supporting a scenario of dynamo action. 
Similarly, the $T_{\bm{uu}}$ plot mostly shows forward transfers as one would expect for a fairly turbulent flow.
Thus, overall, the transfer function plots in this case of non-zero initial velocity, fail to uncover any signatures of reconnection-based inverse transfer. 

Nonetheless, an interesting  feature can be observed in the $T_{\bm{ub}}$ plot. While most of the energy transfers are from low wavenumbers in the kinetic energy reservoir to the high wavenumbers in the magnetic energy reservoir, there is also energy transfer to smaller wavenumbers. For example, the wavenumber $Q=10$ contributes significant energy to $K=7-9$. 
This could be the tail of the small-scale dynamo at low wavenumbers \citep{haugen2004, bhatetal2013}. Then the question which arises is why is there an inverse transfer in decaying turbulence with dynamo effects. In such a system, the eddies which are supercritical to carry out the dynamo action would pertain to the peak in the kinetic spectrum. It can, then, be seen from the \Fig{3dinitvel}, that due to selective decay, this peak shifts to the lower wavenumbers. As the peak in the kinetic spectrum shifts, it could also shift the scales at which the magnetic energy grows, thus leading to an effect that resembles the inverse transfer. A similar effect of flow (which exhibits inverse transfer) dragging the field 
could be the reason for the growth of magnetic energy at $k=1$ as seen in \Fig{2dinitvel} in the 2D simulation, F2D.
A more detailed investigation of the case of non-zero initial velocity field in decaying nonhelical MHD turbulence is left to a future paper.

\section{Discussion and  Conclusions} 
\label{conclusion}

We have investigated the inverse transfer of magnetic energy in the decay of nonhelical MHD
turbulence in 2D and 3D simulations.  We find that the scaling of magnetic energy with time ($\sim
t^{-1}$) and that of power spectrum with wavenumber ($\sim k^{-2}$) is similar between both 2D and 3D
cases (when the initial velocity field is zero). This is suggestive of similar mechanisms being responsible, in both cases, for the inverse
transfer.  In the 2D case, \citet{Munietal2019} have shown that island mergers via magnetic
reconnection are key to understanding formation of larger and larger structures that lead to
inverse transfer. We find that our simulation results support the idea that magnetic reconnection is
responsible for the inverse transfer in 3D nonhelical turbulent systems as well. 

Our investigations have yielded two main results via the study of conserved quantities, timescales, and length scales (via transfer function plots).
In 2D MHD, the ideal invariants include energy and
vector-potential squared. We have provided analytical arguments to show that in a turbulent system,
for large Lundquist numbers, vector-potential squared $P$ is better conserved than magnetic energy
$\EEM$ (the dominant component of enegry in our system) and how, for a decaying system, this can lead to inverse energy transfer. We have calculated the
rate of change of the two ideal invariants from the 2D simulation and shown that indeed $P$ is
better conserved than $\EEM$. Further, we found that this was the case even in the 3D simulations,
indicating that the dynamics in 3D have 2D-like tendencies. This is our first main result.  

Our second main result is that this inverse transfer, both in 2D and in 3D, is due to magnetic reconnection.  
Indeed, on normalizing the time axis by the magnetic reconnection timescale,
we find the evolution curves of the magnetic energry from runs with varying values of Lundquist
numbers collapse on top of each other in both 2D and 3D  (the collapse being better with larger
values of $S$).   
Additionally,  the transfer function plots show clear signatures 
of magnetic reconnection driving the inverse transfer. We find from the $T_{bb}$ plots 
that only those scales either at or above the peak correlation scale, at any given time, 
exhibit inverse transfer as expected from a physical  picture of island (or filament) mergers being dominant at 
a certain scale. The more clinching evidence arises from the $T_{ub}$ plots, 
where it is seen that a set of scales compatible with our understanding of the reconnection process in this system stand out in the transfer of magnetic to kinetic energy. 

From these results, an emergent characteristic of the magnetically dominated 3D system is its tendency to align with the
behaviour observed in 2D.  The overarching question is then what element in 3D renders it with
2D-like behaviour? We think the answer lies in the fact that the system is magnetically dominated. 
The field can provide anisotropy at small-scales i.e. the current sheets can have local guide
fields.  Magnetic reconnection in 3D, when presided by guide field, leads to familiar 2D results
\citep{onofri2004}. Indeed, in another recent study of inverse energy transfer using the reduced-MHD model (which assumes a strong background magnetic field),
\cite{munietal2020} find mergers between magnetic flux tubes driving inverse transfer.   

Returning to the result of $k^{-2}$ slope in the magnetic power spectrum, it has been pointed out
that this corresponds to the theoretical expectation for weak turbulence \citep{axel2015}. However
\citet{Munietal2019} find in their 2D simulations that it corresponds to the presence of thin
current sheets. In accordance with our findings of 2D-like behaviour in 3D, this explanation of
thin current sheets for $k^{-2}$ slope may carry over to 3D as well. \cite{munietal2020} report a $k^{-1.5}$ slope in their reduced-MHD simulations but unlike the case in the simulations here, they also find kinetic energy is not subdominant to the magnetic energy. 

To ascertain whether by making magnetic field subdominant, the inverse transfer in energy
ceases to appear, we performed simulations where the initial velocity was set to a large non-zero value. In the 2D simulation, the system becomes turbulent leading to much faster decay of energy, likely due to anomalous diffusion and there is no significant inverse transfer. In contrast, in the 3D case, the energy decay follows a $t^{-1.4}$ scaling, and we do observe a definitive signature of inverse transfer in the evolution of the magnetic spectrum. Furthermore, the evolution of magnetic energy reveals a dynamo effect which possibly counters the anomalous diffusion, leading to a decay rate that is slower than the one seen in the 2D case. 
On studying the transfer function plots for the 3D simulation, we find that the signature for inverse transfer is surprisingly absent in $T_{\rm bb}$ plot. However, the $T_{\rm ub}$ reveals that there is transfer of energy from the kinetic energy reservoir to magnetic energy to both small and large scales, where the larger portion goes to the small scales.  
This kinetic energy transfer to larger magnetic scales is a possible signature of the tail of small-scale dynamo action at small wavenumbers. This tail can possibly shift further to lower wavenumbers as the peak in kinetic spectrum shifts due to selective decay, leading to an inverse transfer type effect (as seen in the evolving magnetic spectra). 

We have mentioned several astrophysical and cosmological applications to which our results might be relevant in the introduction section. 
In all of the applications mentioned, the astrophysical systems under consideration consist of
highly conducting, large Lundquist number (or magnetic Reynolds number) plasmas.  The range of
Lundquist numbers explored in this paper is limited by the resolution and thus our simulations are
in a regime where Sweet-Parker model for magnetic reconnection is valid. However at higher values of
$S$, the nature of reconnection changes with the onset of the plasmoid instability \citep{nunoetal2007}.
Recent research has revealed that the plasmoid instability renders the magnetic reconnection rate
independent of $S$ for $S\gtrsim 10^4$, with a reconnection rate of $\sim$ 0.01
$V_A$ \citep{bhattacharjeeHuang2009,uzdensky2010,loureiro_2016}.  This would be the timescale to be considered in the astrophysical systems which can be described with the MHD framework. If, instead, the environment under consideration is weakly collisional, the adequate reconnection rate to consider would be faster, on the order of $0.1 V_A$ \citep[e.g.,][]{cassak_2017}

A previous study of this problem had shown that the inverse transfer is weak or altogether absent upon
increasing the magnetic Prandtl number $\Pm$ \citep{reppinbanerjee2017}. This is consistent with the
understanding that magnetic reconnection at higher $\Pm$ becomes increasingly inefficient \citep{park84}. However, it is not clear  if at both higher $S$ and $\Pm$
this trend will continue, as the ensuing plasmoid instability could potentially change it~\citep{loureiroetal2013}. 

In conclusion, we provide a physical understanding to the puzzling and unexpected 3D nonhelical inverse transfer via
analysis of  direct numerical simulations of magnetically dominated, decaying MHD turbulence. We argue that magnetic reconnection is the physical mechanism responsible for the emergence of progressively larger structures. Further, we show that the behavior in the 3D system is intriguingly similar to that in 2D, possibly because of local anisotropy in this system. These results could have important consequences for a wider range of astrophysical applications.

\section*{Acknowledgments}
We thank K. Subramanian for useful feedback on the paper. 
PB and NFL acknowledge support from the NSF-DOE Partnership in Basic Plasma Science and Engineering Award
No. DE-SC0016215. 
MZ and NFL acknowledge support from the NSF CAREER Award No. 1654168.
This project was completed using funding from the European Research Council (ERC) under the European Union’s Horizon 2020 research and innovation programme (grant agreement no. D5S-DLV-786780). The simulations presented in this paper were performed on the MIT-PSFC partition of the Engaging cluster at the MGHPCC facility, funded by DOE Award No. DE-FG02-91-ER54109.

\appendix

\bibliographystyle{mn2e}
\bibliography{invrec3d}

\begin{thebibliography}{}

\bibitem[\protect\citeauthoryear{{Asano} \& {Terasawa}}{{Asano} \&
  {Terasawa}}{2015}]{asanoTerasawa2015}
{Asano} K.,  {Terasawa} T.,  2015, \mnras, 454, 2242

\bibitem[\protect\citeauthoryear{{Baggaley}, {Barenghi} \&
  {Sergeev}}{{Baggaley} et~al.}{2014}]{Baggaleyetal2014}
{Baggaley} A.~W.,  {Barenghi} C.~F.,    {Sergeev} Y.~A.,  2014, \pre, 89,
  013002

\bibitem[\protect\citeauthoryear{{Banerjee} \& {Jedamzik}}{{Banerjee} \&
  {Jedamzik}}{2004}]{banerjeeJedamzik2004}
{Banerjee} R.,  {Jedamzik} K.,  2004, \prd, 70, 123003

\bibitem[\protect\citeauthoryear{{Batchelor}}{{Batchelor}}{1969}]{Batch1969}
{Batchelor} G.~K.,  1969, \pof, 12

\bibitem[\protect\citeauthoryear{{Beck}, {Dolag}, {Lesch} \& {Kronberg}}{{Beck}
  et~al.}{2013}]{becketal2013}
{Beck} A.~M.,  {Dolag} K.,  {Lesch} H.,    {Kronberg} P.~P.,  2013, \mnras,
  435, 3575

\bibitem[\protect\citeauthoryear{{Berera} \& {Linkmann}}{{Berera} \&
  {Linkmann}}{2014}]{bereraLinkmann2014}
{Berera} A.,  {Linkmann} M.,  2014, \pre, 90, 041003

\bibitem[\protect\citeauthoryear{{Bhat} \& {Subramanian}}{{Bhat} \&
  {Subramanian}}{2013}]{bhatetal2013}
{Bhat} P.,  {Subramanian} K.,  2013, \mnras, 429, 2469

\bibitem[\protect\citeauthoryear{{Bhattacharjee}, {Huang}, {Yang} \&
  {Rogers}}{{Bhattacharjee} et~al.}{2009}]{bhattacharjeeHuang2009}
{Bhattacharjee} A.,  {Huang} Y.-M.,  {Yang} H.,    {Rogers} B.,  2009, \pop,
  16, 112102

\bibitem[\protect\citeauthoryear{{Biferale}, {Musacchio} \&
  {Toschi}}{{Biferale} et~al.}{2012}]{Biferaleetal2012}
{Biferale} L.,  {Musacchio} S.,    {Toschi} F.,  2012, \prl, 108, 164501

\bibitem[\protect\citeauthoryear{{Biskamp}}{{Biskamp}}{2003}]{Biskamp2003}
{Biskamp} D.,  2003, {Magnetohydrodynamic Turbulence}

\bibitem[\protect\citeauthoryear{{Biskamp} \& {Welter}}{{Biskamp} \&
  {Welter}}{1989}]{BW1989}
{Biskamp} D.,  {Welter} H.,  1989, \pofb, 1, 1964

\bibitem[\protect\citeauthoryear{{Blandford}, {Yuan}, {Hoshino} \&
  {Sironi}}{{Blandford} et~al.}{2017}]{blandfordetal2017}
{Blandford} R.,  {Yuan} Y.,  {Hoshino} M.,    {Sironi} L.,  2017, \ssr, 207,
  291

\bibitem[\protect\citeauthoryear{Brandenburg, Kahniashvili \&
  Tevzadze}{Brandenburg et~al.}{2015}]{axel2015}
Brandenburg A.,  Kahniashvili T.,    Tevzadze A.~G.,  2015, \prl, 114, 075001

\bibitem[\protect\citeauthoryear{{Brandenburg} \&
  {K{\"a}pyl{\"a}}}{{Brandenburg} \& {K{\"a}pyl{\"a}}}{2007}]{axelkapyla2007}
{Brandenburg} A.,  {K{\"a}pyl{\"a}} P.~J.,  2007, \njp, 9, 305

\bibitem[\protect\citeauthoryear{{Brandenburg} \& {Subramanian}}{{Brandenburg}
  \& {Subramanian}}{2005}]{BS05}
{Brandenburg} A.,  {Subramanian} K.,  2005, \physrep, 417, 1

\bibitem[\protect\citeauthoryear{Burgers}{Burgers}{1948}]{burgers_1948}
Burgers J.~M.,  1948, in , Vol.~1, Advances in applied mechanics.
Elsevier, pp 171--199

\bibitem[\protect\citeauthoryear{Cassak, Liu \& Shay}{Cassak
  et~al.}{2017}]{cassak_2017}
Cassak P.~A.,  Liu Y.-H.,    Shay M.,  2017, \jpp, 83, 715830501

\bibitem[\protect\citeauthoryear{{Christensson}, {Hindmarsh} \&
  {Brandenburg}}{{Christensson} et~al.}{2001}]{CHB2001}
{Christensson} M.,  {Hindmarsh} M.,    {Brandenburg} A.,  2001, \pre, 64,
  056405

\bibitem[\protect\citeauthoryear{{Davidson}}{{Davidson}}{2000}]{davidson}
{Davidson} P.~A.,  2000, \jott, 1, 6

\bibitem[\protect\citeauthoryear{{Davidson}}{{Davidson}}{2004}]{Davidson2004}
{Davidson} P.~A.,  2004, {Turbulence : an introduction for scientists and
  engineers}.
Oxford University Press

\bibitem[\protect\citeauthoryear{{Falkovich} \& {Sreenivasan}}{{Falkovich} \&
  {Sreenivasan}}{2006}]{FS2006}
{Falkovich} G.,  {Sreenivasan} K.~R.,  2006, \ptod, 59, 43

\bibitem[\protect\citeauthoryear{{Frisch}}{{Frisch}}{1995}]{Frisch1995}
{Frisch} U.,  1995, {Turbulence. The legacy of A. N. Kolmogorov}.
Cambridge University Press

\bibitem[\protect\citeauthoryear{{Fyfe} \& {Montgomery}}{{Fyfe} \&
  {Montgomery}}{1976}]{FM76}
{Fyfe} D.,  {Montgomery} D.,  1976, \jpp, 16, 181

\bibitem[\protect\citeauthoryear{{Gao}, {Xu} \& {Law}}{{Gao}
  et~al.}{2015}]{gaoetal2015}
{Gao} Y.,  {Xu} H.,    {Law} C.~K.,  2015, \apj, 799, 227

\bibitem[\protect\citeauthoryear{{Grete}, {O'Shea}, {Beckwith}, {Schmidt} \&
  {Christlieb}}{{Grete} et~al.}{2017}]{Greteetal17}
{Grete} P.,  {O'Shea} B.~W.,  {Beckwith} K.,  {Schmidt} W.,    {Christlieb} A.,
   2017, \pop, 24, 092311

\bibitem[\protect\citeauthoryear{{Haugen}, {Brandenburg} \& {Dobler}}{{Haugen}
  et~al.}{2004}]{haugen2004}
{Haugen} N.~E.,  {Brandenburg} A.,    {Dobler} W.,  2004, \pre, 70, 016308

\bibitem[\protect\citeauthoryear{{Kadomtsev} \& {Pogutse}}{{Kadomtsev} \&
  {Pogutse}}{1974}]{kadomtsev_1974}
{Kadomtsev} B.~B.,  {Pogutse} O.~P.,  1974, \zhetp, 38, 283

\bibitem[\protect\citeauthoryear{{Kraichnan}}{{Kraichnan}}{1967}]{Kraichnan1967}
{Kraichnan} R.~H.,  1967, \pof, 10, 1417

\bibitem[\protect\citeauthoryear{{Loureiro}, {Schekochihin} \&
  {Cowley}}{{Loureiro} et~al.}{2007}]{nunoetal2007}
{Loureiro} N.~F.,  {Schekochihin} A.~A.,    {Cowley} S.~C.,  2007, \pop, 14,
  100703

\bibitem[\protect\citeauthoryear{{Loureiro}, {Schekochihin} \&
  {Uzdensky}}{{Loureiro} et~al.}{2013}]{loureiroetal2013}
{Loureiro} N.~F.,  {Schekochihin} A.~A.,    {Uzdensky} D.~A.,  2013, \pre, 87,
  013102

\bibitem[\protect\citeauthoryear{{Loureiro} \& {Uzdensky}}{{Loureiro} \&
  {Uzdensky}}{2016}]{loureiro_2016}
{Loureiro} N.~F.,  {Uzdensky} D.~A.,  2016, PPCF, 58, 014021

\bibitem[\protect\citeauthoryear{{Mac Low}, {Klessen}, {Burkert} \&
  {Smith}}{{Mac Low} et~al.}{1998}]{maclowetal1998}
{Mac Low} M.-M.,  {Klessen} R.~S.,  {Burkert} A.,    {Smith} M.~D.,  1998,
  \prl, 80, 2754

\bibitem[\protect\citeauthoryear{{Moffatt}}{{Moffatt}}{1978}]{moffatt}
{Moffatt} H.~K.,  1978, {Magnetic field generation in electrically conducting
  fluids}.
Cambridge University Press

\bibitem[\protect\citeauthoryear{{Nazarenko}}{{Nazarenko}}{2011}]{Naz2011}
{Nazarenko} S.,  ed. 2011, {Wave Turbulence} Vol.~825 of Lecture Notes in
  Physics, Berlin Springer Verlag

\bibitem[\protect\citeauthoryear{{Onofri}, {Primavera}, {Malara} \&
  {Veltri}}{{Onofri} et~al.}{2004}]{onofri2004}
{Onofri} M.,  {Primavera} L.,  {Malara} F.,    {Veltri} P.,  2004, \pop, 11,
  4837

\bibitem[\protect\citeauthoryear{{Park}, {Monticello} \& {White}}{{Park}
  et~al.}{1984}]{park84}
{Park} W.,  {Monticello} D.~A.,    {White} R.~B.,  1984, \pof, 27, 137

\bibitem[\protect\citeauthoryear{{Parker}}{{Parker}}{1957}]{Parker1957}
{Parker} E.~N.,  1957, \jgr, 62, 509

\bibitem[\protect\citeauthoryear{{Pouquet}}{{Pouquet}}{1978}]{pouquet_1978}
{Pouquet} A.,  1978, \jfm, 88, 1

\bibitem[\protect\citeauthoryear{{Pouquet}, {Frisch} \& {Leorat}}{{Pouquet}
  et~al.}{1976}]{PFL76}
{Pouquet} A.,  {Frisch} U.,    {Leorat} J.,  1976, \jfm, 77, 321

\bibitem[\protect\citeauthoryear{{Pouquet}, {Rosenberg}, {Stawarz} \&
  {Marino}}{{Pouquet} et~al.}{2019}]{Pouquetetal2019}
{Pouquet} A.,  {Rosenberg} D.,  {Stawarz} J.~E.,    {Marino} R.,  2019, Earth
  and Space Science, 6, 351

\bibitem[\protect\citeauthoryear{{Reppin} \& {Banerjee}}{{Reppin} \&
  {Banerjee}}{2017}]{reppinbanerjee2017}
{Reppin} J.,  {Banerjee} R.,  2017, \pre, 96, 053105

\bibitem[\protect\citeauthoryear{{Samtaney}, {Loureiro}, {Uzdensky},
  {Schekochihin} \& {Cowley}}{{Samtaney} et~al.}{2009}]{samtaney_2009}
{Samtaney} R.,  {Loureiro} N.~F.,  {Uzdensky} D.~A.,  {Schekochihin} A.~A.,
  {Cowley} S.~C.,  2009, \prl, 103, 105004

\bibitem[\protect\citeauthoryear{{Schekochihin}, {Cowley}, {Dorland},
  {Hammett}, {Howes}, {Quataert} \& {Tatsuno}}{{Schekochihin}
  et~al.}{2009}]{alex_2009}
{Schekochihin} A.~A.,  {Cowley} S.~C.,  {Dorland} W.,  {Hammett} G.~W.,
  {Howes} G.~G.,  {Quataert} E.,    {Tatsuno} T.,  2009, \apjs, 182, 310

\bibitem[\protect\citeauthoryear{{Sethi} \& {Subramanian}}{{Sethi} \&
  {Subramanian}}{2005}]{sethikandu2005}
{Sethi} S.~K.,  {Subramanian} K.,  2005, \mnras, 356, 778

\bibitem[\protect\citeauthoryear{{Skrbek} \& {Sreenivasan}}{{Skrbek} \&
  {Sreenivasan}}{2012}]{SS2012}
{Skrbek} L.,  {Sreenivasan} K.~R.,  2012, \pof, 24, 011301

\bibitem[\protect\citeauthoryear{{Strauss}}{{Strauss}}{1976}]{strauss_1976}
{Strauss} H.~R.,  1976, \pof, 19, 134

\bibitem[\protect\citeauthoryear{{Subramanian}}{{Subramanian}}{2016}]{kandu2016}
{Subramanian} K.,  2016, \rpp, 79, 076901

\bibitem[\protect\citeauthoryear{{Subramanian}, {Shukurov} \&
  {Haugen}}{{Subramanian} et~al.}{2006}]{sss2006}
{Subramanian} K.,  {Shukurov} A.,    {Haugen} N. E.~L.,  2006, \mnras, 366,
  1437

\bibitem[\protect\citeauthoryear{{Sur}}{{Sur}}{2019}]{sur2019}
{Sur} S.,  2019, \mnras, 488, 3439

\bibitem[\protect\citeauthoryear{{Sweet}}{{Sweet}}{1958}]{Sweet1958}
{Sweet} P.~A.,  1958, in {Lehnert} B.,  ed., Electromagnetic Phenomena in
  Cosmical Physics Vol.~6 of IAU Symposium.
p.~123

\bibitem[\protect\citeauthoryear{{Uzdensky}, {Loureiro} \&
  {Schekochihin}}{{Uzdensky} et~al.}{2010}]{uzdensky2010}
{Uzdensky} D.~A.,  {Loureiro} N.~F.,    {Schekochihin} A.~A.,  2010, \prl, 105,
  235002

\bibitem[\protect\citeauthoryear{{Yakhot} \& {Pelz}}{{Yakhot} \&
  {Pelz}}{1987}]{YakhotPelz87}
{Yakhot} V.,  {Pelz} R.,  1987, \pof, 30, 1272

\bibitem[\protect\citeauthoryear{{Zeldovich}}{{Zeldovich}}{1957}]{Z57}
{Zeldovich} r.~B.,  1957, \zhetp, 4, 460

\bibitem[\protect\citeauthoryear{{Zhou}, {Bhat}, {Loureiro} \&
  {Uzdensky}}{{Zhou} et~al.}{2019}]{Munietal2019}
{Zhou} M.,  {Bhat} P.,  {Loureiro} N.~F.,    {Uzdensky} D.~A.,  2019, \prr, 1,
  012004

\bibitem[\protect\citeauthoryear{Zhou, Loureiro \& Uzdensky}{Zhou
  et~al.}{2020}]{munietal2020}
Zhou M.,  Loureiro N.~F.,    Uzdensky D.~A.,  2020, \jpp, 86, 535860401

\bibitem[\protect\citeauthoryear{Zrake}{Zrake}{2014}]{zrake2014}
Zrake J.,  2014, ApJL, 794, L26

\bibitem[\protect\citeauthoryear{{Zrake}}{{Zrake}}{2016}]{zrake2016}
{Zrake} J.,  2016, \apj, 823, 39

\end{thebibliography}

\label{lastpage}

\end{document}